\documentclass[conference,compsoc]{IEEEtran}

\usepackage{url}
\usepackage{tikz}
\usepackage{amsmath}
\usepackage{amsthm}
\usepackage{amssymb}
\usepackage{todonotes}
\usepackage{algorithm}
\usepackage{algpseudocode}
\usepackage{listings}
\usepackage{makecell}
\usepackage{multirow}
\usepackage[hidelinks]{hyperref}

\usepackage{tikz}
\usepackage{booktabs}
\usetikzlibrary{matrix}
\theoremstyle{plain}
\newtheorem{theorem}{Theorem}
\newtheorem{collorary}{Collorary}
\newtheorem{assumption}{Assumption}
\newtheorem{lemma}{Lemma}
\theoremstyle{definition}
\newtheorem{definition}{Definition}
\newtheorem{example}{Example}

\algrenewcommand\algorithmicindent{1em}
\ifCLASSOPTIONcompsoc
  \usepackage[nocompress]{cite}
\else
  \usepackage{cite}
\fi

\lstdefinelanguage
   [x64]{Assembler}     
   [x86masm]{Assembler} 
   {morekeywords={CDQE,CQO,CMPSQ,CMPXCHG16B,JRCXZ,LODSQ,MOVSXD, %
       POPFQ,PUSHFQ,SCASQ,STOSQ,IRETQ,RDTSCP,SWAPGS, %
       jmpq,nopw,
                  rax,rdx,rcx,rbx,rsi,rdi,rsp,rbp, %
                  r8,r8d,r8w,r8b,r9,r9d,r9w,r9b, %
                  r10,r10d,r10w,r10b,r11,r11d,r11w,r11b, %
                  r12,r12d,r12w,r12b,r13,r13d,r13w,r13b, %
                  r14,r14d,r14w,r14b,r15,r15d,r15w,r15b,rip},
     columns=fixed,
     basicstyle=\small\ttfamily\upshape,
   } 

\newcommand{\instr}{\mathit{instr}}
\newcommand{\instrEnd}{\mathit{end}}
\newcommand{\instrStart}{\mathit{start}}

\newcommand{\Code}{\mathit{Code}}
\newcommand{\Data}{\mathit{Data}}
\newcommand{\CodeBlock}{\mathit{codeBlock}}
\newcommand{\DataBlock}{\mathit{dataBlock}}

\newcommand{\Block}{\mathit{block}}

\newcommand{\BlockSet}{\mathit{Blocks}}
\newcommand{\CandidateBlocks}{\mathit{CBlocks}}

\newcommand{\blockSize}{\mathit{size}}

\newcommand{\bweight}{\mathit{W}}
\newcommand{\sweight}{\mathit{SW}}

\newcommand{\heuristicMatch}{\mathit{\#}}
\newcommand{\heuristicSet}{H}
\newcommand{\heuristicSets}{\heuristicSet_s}
\newcommand{\heuristicSetp}{\heuristicSet_p}

\newcommand{\BlockInsn}{\mathit{Insns}}
\newcommand{\TrueCode}{\mathit{TCode}}
\newcommand{\Ignored}{\mathit{Ignored}}
\newcommand{\BlockOf}{\mathit{BlockOf}}

\newcommand{\TrueBlocks}{\mathit{TBlocks}}
\newcommand{\FalseBlocks}{\mathit{FBlocks}}

\newcommand{\opt}{\mathit{opt}}
\newcommand{\symopt}{\mathit{symopt}}

\newcommand{\SSchedWeight}{weight}
\newcommand{\SSchedScore}{score}

\newcommand{\score}{\mathit{score}}
\newcommand{\updateScore}{\mathit{uScore}}

\newcommand{\bestSchedule}{\mathit{bestSched}}

\newcommand{\simplifyS}{\mathit{simplifyS}}
\newcommand{\simplifyC}{\mathit{simplifyC}}

\newcommand{\constraintSet}{\mathit{Cs}}

\newcommand{\slackSet}{\mathit{SKs}}

\newcommand{\ArmISA}{arm32}
\newcommand{\AarchISA}{aarch64}

\newif\ifanonymous
\anonymousfalse

\newcommand{\anonymizedUrl}[1]{
\ifanonymous
  The URL has been anonymized for the submission.
\else
\url{#1}
\fi
}

\newcommand{\tool}{Ddisasm-WIS}

\newcommand\copyrighttext{%
	\footnotesize \textcopyright 2025 IEEE. Personal use of this material is 
	permitted. Permission from IEEE must be
	obtained for all other uses, in any current or future media, including
	reprinting/republishing this material for advertising or promotional 
	purposes, creating new
	collective works, for resale or redistribution to servers or lists, or 
	reuse of any copyrighted
	component of this work in other works.}
\newcommand\copyrightnotice{%
	\begin{tikzpicture}[remember picture,overlay]
		\node[anchor=south,yshift=10pt] at (current page.south) 
		{\fbox{\parbox{\dimexpr\textwidth-\fboxsep-\fboxrule\relax}{\copyrighttext}}};
	\end{tikzpicture}%
}

\begin{document}

\date{}

\title{Disassembly as Weighted Interval Scheduling with Learned Weights}

\ifanonymous
 \author{}
\else
%
\author{\IEEEauthorblockN{
  Antonio Flores-Montoya\IEEEauthorrefmark{1},
  Junghee Lim \IEEEauthorrefmark{1},
  Adam Seitz \IEEEauthorrefmark{1}, 
  Akshay Sood \IEEEauthorrefmark{1},
  Edward Raff \IEEEauthorrefmark{2} and
  James Holt \IEEEauthorrefmark{3}}
\IEEEauthorblockA{\IEEEauthorrefmark{1}GrammaTech Inc.}
\IEEEauthorblockA{\IEEEauthorrefmark{2}Booz Allen Hamilton}
\IEEEauthorblockA{\IEEEauthorrefmark{3}Laboratory for Physical Sciences}
}
\fi 
\maketitle
\copyrightnotice

\begin{abstract}
Disassembly is the first step of a variety of binary analysis and
transformation techniques, such as reverse engineering, or binary rewriting.
Recent disassembly approaches consist of three phases:
an exploration phase, that overapproximates the binary's code;
an analysis phase, that assigns weights to candidate instructions or basic blocks;
and a conflict resolution phase, that downselects the final set of instructions.
We present a disassembly algorithm that generalizes this pattern for a 
wide range of architectures, namely x86, x64, arm32, and \AarchISA{}.
Our algorithm presents a novel conflict resolution method that reduces disassembly
to weighted interval scheduling.
Additionally, we present a weight assignment algorithm that allows us to 
learn optimal weights for the various disassembly heuristics in the analysis phase.
Learned weights outperform manually tuned weights in most cases while reducing the number
of necessary heuristics by 40\% (by setting their weights to zero).
Our implementation, built on top of Ddisasm, outperforms state-of-the-art
disassemblers in several metrics and achieves the largest proportion of
perfectly disassembled binaries by a wide margin in all evaluated datasets.
\end{abstract}

\section{Introduction}
\label{sec:intro}

Disassembly is the process of recovering assembly instructions from
binary code. This amounts to deciding, for each byte in the binary
program, whether it belongs to an instruction or should be
interpreted as data.  Disassembly is an essential first step of a
variety of binary analysis and transformation techniques, including
reverse engineering, binary rewriting, or
vulnerability discovery.

Traditional approaches rely on linear sweep or recursive
disassembly.  Linear sweep starts decoding instructions at the
beginning of the code section and proceeds sequentially. This approach
results in errors when data is interleaved with code. Recursive
disassembly recovers the assembly code by traversing the control flow
of the program, but is disadvantaged in resolving indirect branches
limiting its coverage.

Modern reverse engineering
frameworks~\cite{ghidra,ida,binja,hopper,angr,BAP,radare} often use a
combination of both linear and recursive disassembly, relying on heuristics to find additional
code or make decisions among conflicting code blocks~\cite{sok-x86}.
These techniques and heuristics are often Instruction Set Architecture
(ISA) specific.
Several of the more recent techniques~\cite{probabilistic,ddisasm,d-arm,accuratedisassembly} can be thought of as having the following steps:
\begin{enumerate}
\item An exploration phase in which a superset of all the potential instructions or data blocks is collected.
\item An analysis phase in which static analysis is performed to gather evidence for and against
candidate instructions or data blocks, usually in the form of
dependencies between candidates or weights assigned to
each candidate.
\item A conflict resolution phase in which the evidence is aggregated to reach a final decision.
\end{enumerate}

For example, D-ARM~\cite{d-arm} proposes an approach for ARM in
which the exploration phase considers all possibilities, i.e. every
aligned address can be ARM code, Thumb code, or data; and expresses the
conflict resolution problem as \emph{maximum weight independent set}
(MWIS) optimization problem~\cite{MWIS}.  Note that MWIS is
NP-hard and D-ARM's implementation needs to resort to a greedy approximation.

Ddisasm~\cite{ddisasm}, D-ARM~\cite{d-arm}, and other comparable approaches
suffer what we will refer to as
the \emph{weight assignment problem}. Given a set of analyses,
heuristics and a conflict resolution method, what are the optimal
weights that should be assigned to each of the heuristics to maximize
performance? Are all heuristics necessary? Previous approaches rely on hand-tuned weights~\cite{ddisasm}
or ISA-specific statistical studies combined and ad-hoc aggregation methods~\cite{accuratedisassembly}.

In our work, we present a multi-ISA (with support for x86, x64, \ArmISA{}, and \AarchISA{}
binaries) disassembly approach that follows the three generic steps
described above, but presents two key innovations. First, it relies on
a tractable yet powerful conflict resolution algorithm. Our
disassembly algorithm expresses the conflict resolution problem as a
Weighted Interval Scheduling (WIS) problem, for which there is a
complete and efficient dynamic programming algorithm~\cite{kleinberg}.
In this setting, each code block or data block is considered a
``task'' to be scheduled with the restriction that tasks cannot
overlap with each other.  The resulting disassembly corresponds to the
optimal schedule (the schedule with maximum aggregated weight).

Second, given this conflict resolution algorithm, we provide a solution to the
weight assignment problem by encoding the inference of optimal
heuristic weights with respect to a set of binaries
with known ground truth as a linear programming (LP) problem
with soft constraints. This allows us to obtain optimal heuristic
weights using off-the-self solvers (like Pulp~\cite{pulp})
even in cases when no perfect solution exists.

We build our implementation on top of Ddisasm~\cite{ddisasm} and
evaluate it in comparison with other state-of-the-art tools on several
ISAs. We perform this evaluation while accounting for the limitations
of the existing approaches to gather ground
truth~\cite{gt-is-not-easy,pangines-gt}.  In addition, we evaluate
individual aspects of the algorithm: the completeness of
the exploration phase and the generalization properties of the weight
inference algorithm. Finally, we discuss some insights about the learned
heuristics weights.

In summary, our contributions are the following:
\begin{enumerate}
\item A multi-ISA disassembly algorithm that performs conflict resolution
among candidate instructions and data blocks by solving a weighted interval
scheduling (WIS) optimization problem (Sect.~\ref{sec:algorithm}).
\item A weight assignment algorithm that optimizes the weights
assigned to the heuristics used in the disassembly algorithm based on
a corpus of ground-truth annotated binaries
(Sect.~\ref{sec:weight_assignment}).  This algorithm supports
incomplete annotations to account for the limitations in ground truth
extraction techniques.
\item An extensive experimental evaluation of the disassembly algorithm
and its corresponding weight assignment algorithm that includes a
large collection of \ArmISA{}, \AarchISA{}, x86, and x64 stripped binaries
compiled with various compilers, optimization levels, and binary
formats (ELF and PE) including the datasets from
~\cite{pangines-gt,gt-is-not-easy,arm-evaluation,assemblage}
(Sect.~\ref{sec:experiments}).
\end{enumerate}

\section{Static Disassembly Instruction Recovery}
\label{sec:problem}
In this section, we define static disassembly and
state some simplifying assumptions that frame the problem.

\subsection{Basic Definitions}
\label{section:basic-definitions}
Informally, we refer to disassembly as the process of recovering
assembly instructions from binary code, which amounts to deciding the
location and decode mode of the assembly instructions in the binary's
executable sections.

We restrict ourselves to \emph{static} disassembly, i.e., we only
consider instructions present in the binary file. In other words, we do
not consider dynamically generated or self-modifying code.

Let us first define decode mode and instruction decoding.
The default decode mode for any ISA will be referred to as $A$, 
and \ArmISA{}'s second decode mode (Thumb, as opposed to ARM) will be referred to as $T$.
Instruction decoding maps a sequence of bytes and a decode mode to an assembly instruction.
As such, instruction decoding is deterministic, i.e. an address and a
decode mode uniquely determine the instruction located at that address and
its size.
  Thus, given a binary
section spanning addresses $[s,e)$, let $d \in \{ A, T \}$ be a decode
  mode and $a\in [s,e)$ an address within the binary section. We
    define a partial function $\instr(a,d)$ that defines the assembly
    instruction starting at address $a$ and using the decode mode
    $d$\footnote{$\instr(a,d)$ is partial because there might be
    combinations of decode modes and addresses that do not define a
    valid instruction in the given ISA.}.  The function
    $\instrEnd(\instr(a,d))$ denotes the end address of the
    instruction $\instr(a,d)$.

Given this characterization of instructions, we define a binary's
$\Code$ as a set of instructions based on the binary program
semantics.

\begin{definition}[Code]
An instruction $\instr(a,d)$ is code $\instr(a,d)\in \Code$ if there exists a program
trace that contains the instruction $\instr(a,d)$.
\end{definition}

Under this definition, unreachable instructions (dead code)
are not considered code, and determining code accurately is not
decidable (since it is connected to reachability).
Engel et al.~\cite{disassemblyDecidability}'s sound disassembly decidability theorems
allow overapproximating $\Code$ (which is indeed decidable in our setting), whereas our goal is
to recover it precisely.

This definition also accounts for the challenges in collecting ground
truth for disassembly~\cite{pangines-gt,gt-is-not-easy}. Developers may encode
instructions meant to be executed as data, which confuses
compiler-based ground truth generation.

Similarly, we can define data in a binary program based on its dynamic behavior.
\begin{definition}[Data]
A program address $a$ is \emph{data}, $a\in \Data$, if there exists a
program execution trace that reads or writes that memory location.
\end{definition}

Instruction recovery aims to infer $\Code$.
In general, $\Code$ and $\Data$ are neither complementary nor mutually exclusive.  There
can be regions of a binary that are neither code (these regions are
never executed) nor data (those bytes are never read or written). We
call such regions \emph{padding} since they are introduced by
compilers to ensure alignment.

\begin{example}
  The assembly listing in Figure~\ref{fig:padding} contains padding in
  the address range $[94c67-94c70)$.  Those bytes are never executed\footnote{Actually proving
      that those bytes are never executed would be very challenging but we
      are fairly confident in this particular case. The execution cannot fall through $94c67$
      and we can find jumps that target $94c70$ directly.}
    nor are they read by any instructions.  Some disassemblers will
    interpret them as nops (as in the figure) or as data but they are
    neither according to our definition.
    
\end{example}
\begin{figure}
\begin{lstlisting}
   94c40: je     94c43
   
   94c42: lock cmpxchg %rcx,0x13c6dd(%rip)       
   94c4b: cmp    %rdx,%rax
   94c4e: je     94cfc
   94c54: mov    0x13c6cd(%rip),%rdx       
   94c5b: mov    0x14380e(%rip),%rax    
   94c62: jmpq   94ad4

   94c67: nopw   0x0(%rax,%rax,1)
   94c70: mov    0x870(%r12),%r12
\end{lstlisting}
\caption{x64 assembly snippet extracted from \texttt{glibc-2.36}.
  The example contains both padding in the address range
$[94c67-94c70)$) and a prefix-enclosed instruction at address $94c43$. }
\label{fig:padding}
\end{figure}

A binary region can also be both code and data. The instructions in
that region are executed and also read or written during the lifetime
of the program.

\subsection{Assumptions}
\label{sec:assumptions}

In the previous subsection, we have provided a general definition of
$\Code$ and the desired output of the disassembly
process. Next, we explicitly state two simplifying assumptions.
In general, we find that while padding is extremely common in compiler-generated binaries,
binary regions that are both code and data are uncommon.
Thus, our disassembly algorithm rests on the following assumption:
\begin{assumption}
  $\Code$ and $\Data$ are mutually exclusive.
  \label{assumption:code-data}
\end{assumption}
Thus, our algorithm will approximate $\Data$ to better determine $\Code$.

Our second assumption concerns overlapping instructions. Overlapping
instructions are uncommon, but present in certain notable examples. In
particular, we are aware of certain instances of overlapping
instructions in glibc (e.g., see Figure~\ref{fig:padding}). These overlapping instructions follow a very
specific pattern in which an instruction can be executed with and
without a prefix.  We call these cases \emph{prefix-enclosed
instructions}.

\begin{example}
  \label{example:prefix-enclosed}
  The assembly in Figure~\ref{fig:padding} contains a prefix-enclosed
  instruction at address \texttt{94c43}. The conditional jump \lstinline!je 94c43!
  will fall through to address \lstinline!94c42! or jump
  to address \lstinline!94c43! depending on the condition flag.
  This will result in the execution of
  \lstinline!lock cmpxchg %rcx,0x13c6dd(%rip)! or
  \lstinline!cmpxchg %rcx,0x13c6dd(%rip)!, respectively.
\end{example}

\begin{assumption}
All instructions in $\Code$ are non-overlapping except
for prefix-enclosed instructions.
\label{assumption:instruction-overlap}
\end{assumption}

Note that most disassembly algorithms assume code and data are
mutually exclusive and non-overlapping
instructions~\cite{ddisasm,probabilistic,accuratedisassembly,d-arm}
though these assumptions are often not explicitly stated.

\section{Disassembly Algorithm}
\label{sec:algorithm}

\subsection{Overview}

Our disassembly algorithm relies on the concept of code blocks and
data blocks.  These are sequences of consecutive instructions or
addresses containing data that are treated as a single entity.

\begin{definition}[Code Block]
\label{def:code-block}
A code block is a sequence of contiguous instructions with
the same decode mode.
Let $[s,e)$ be an address range and $d$ be a decode mode, we
  can define a code block $\CodeBlock(s,e,d)$ as a sequence:
  $$[\instr(s_1,d),\instr(s_2,d), \ldots, \instr(s_n,d)]$$
    such that $s_1=s$,  $end(\instr(s_i,d))=s_{i+1}$ for all $1\leq i < n$
    and $end(\instr(s_n,d))=e$.
    Let $b$ be a code block, $\BlockInsn(b)$ denotes the set of all the instructions in $b$.
    Conversely, $\BlockOf(\instr(s,d))$ defines the inverse mapping, i.e. the set of blocks that contain
    instruction $\instr(s,d)$.
\end{definition}

\begin{definition}[Data Block]
  A data block  $\DataBlock(s,e)$,
    is a sequence of contiguous addresses in
    the range $[s,e)$.
\end{definition}

We define a unified $\Block$ notation to represent both code and data blocks.
For that purpose we define a special ``data'' (denoted $D$) decode mode: i.e.,
\begin{description}
  \item $\Block(s,e,D) = \DataBlock(s,e)$
  \item $\Block(s,e,d) = \CodeBlock(s,e,d)$
\end{description}

Given this definition, the result of the disassembly
algorithm is a set of blocks $\mathit{block}([s,e),d)\in \BlockSet$
  which can be mapped to $\Code$ and $\Data$ sets
  as defined in Sect.~\ref{section:basic-definitions}.

The disassembly algorithm has three phases:
\begin{enumerate}
\item \textbf{Candidate Generation}: The candidate generation phase generates a set
  of candidate blocks $\CandidateBlocks$ in which the code blocks should
  be an overapproximation of the final code blocks. It does so by performing
  a traversal of the binary that combines linear sweep and recursive disassembly.
  
\item \textbf{Block Weight Assignment}: The block weight assignment phase performs a series of analyses and
  assigns a weight to each of the candidate blocks based on heuristics.
  That is, it defines a function from candidate blocks to integers $\bweight:
  \CandidateBlocks \rightarrow \mathbb{Z}$ where
  $\bweight(\mathit{block}([s,e),d))$ is the weight
    associated to the candidate block $\Block(s,e,d)$.

\item \textbf{Conflict-Resolution}: The conflict resolution phase selects a
      subset of blocks $\BlockSet$ from $\CandidateBlocks$ ($\BlockSet\subseteq\CandidateBlocks$) based on their
      weight and such that there are no overlaps among the
      selected blocks.
      
\end{enumerate}

Grouping instructions and data bytes into blocks serves two purposes.
(1) It decreases the number of overall candidates for improved
performance, and (2) it provides a coarser granularity for the analysis
and conflict resolution phase.
Heuristics can consider the likelihood of sequences of instructions rather than
just individual instructions.
This is particularly relevant for x86/x64, which has a dense instruction set
and unaligned and variable-size instructions.
For example, let us consider a \texttt{libc} x64 implementation. Its code section is 1.1MB and its superset
disassembly contains 1,128,485 instructions with an average size of 3.2 bytes per instruction. If we generate candidates
for each instruction and data byte, we will have approximately 2M candidates and 5M overlaps.
In contrast, our candidate generation algorithm produces 100,820 candidates and only 5430 overlaps.

The following subsections describe each of the disassembly phases in detail.

\subsection{Candidate Generation}
\label{sec:candidate-generation}

The candidate generation phase generates a set
of candidate blocks $\CandidateBlocks$
and its goal is to ensure that the set is an overapproximation of the final blocks.

Our candidate generation algorithm extends Ddisasm's original
algorithm~\cite{ddisasm}. Ddisasm is implemented in Datalog, and so is
our algorithm.  Datalog can easily implement code traversals using
recursive rules and pattern-matching rules for detecting code
patterns (e.g. jump table patterns).

In Sect.~\ref{sec:original-traversal} we
summarize the original algorithm,
and Sect.~\ref{sec:extended-traversal} describes how this algorithm
has been extended.
Sect.~\ref{sec:candidate-errors} formally
defines the conditions that ensure that our candidate set $\CandidateBlocks$ is overapproximating
and characterizes possible failures during candidate generation.

\subsubsection{Ddisasm's original algorithm}
\label{sec:original-traversal}
The starting point of Ddisasm's original disassembly algorithm is the
superset of all instructions, i.e. it computes $\instr(s,d)$ for every
address $s$ and decode mode $d$.

Ddisasm  performs two traversals over this representation: a backward
traversal conservatively discards instructions that lead (fallthrough or jump)
to invalid locations, and a forward traversal generates candidate code
blocks on the remaining instructions.

The forward traversal combines recursive and linear disassembly. The
traversal of a binary program starts from an initial set of addresses, which is
aggressively computed by considering the entry point of the program,
symbols, exception information, and any sequence of bytes that could
be interpreted as an address anywhere in the binary.

From those starting points, the traversal follows the control flow of
the program. Whenever it encounters a jump or a call, it generates new
starting points both at the jump/call target (if it is direct)
and immediately after the jump. Starting points are added even after unconditional
jumps or calls that are known not to return. That constitutes the
``linear'' component of the traversal.  This traversal does not
attempt to resolve indirect jumps or calls.

A post-processing phase then splits code blocks that have common suffixes,
which ensures that code blocks have the following properties:
  \begin{enumerate}
  \item Candidate code blocks do not share instructions, i.e.,
    each instruction belongs to at most one candidate
    block.
  \item Each instruction within a code block must fall through into the next
    one in the block. This means that instructions that affect the control flow (e.g. jumps
    or function calls)
    or might stop the execution (e.g. \texttt{hlt}) can only be at the end of code blocks.
  \end{enumerate}

\subsubsection{Extended Traversal}
\label{sec:extended-traversal}

The original traversal was designed for Linux x64 binaries where data
and code interleaving are rare~\cite{andriesse}.
Our extended traversal is designed to better support Windows x64, Linux \ArmISA{},
and Linux \AarchISA{} binaries. This extended support requires (1) the incorporation
of multiple decode modes for \ArmISA{} (2) more exhaustive handling of data
and code interleavings. \ArmISA{} binaries make extensive use of literal
pools~\cite{d-arm}, and Windows binaries (compiled with Visual Studio) often have jump tables
in the code section.

For better handling of code and data interleavings, our extended forward traversal
(1) generates candidate data blocks, and (2) extends its linear traversal.

\paragraph{\textbf{Candidate Data block generation}}
The goal of data blocks is to ``compete'' against code blocks and help
us discard spurious code blocks (relying on
Assumption~\ref{assumption:code-data}).  Candidate data blocks are
created whenever the traversal encounters code that accesses (reads or
writes) memory locations in the code sections.  We have specialized
rules for creating candidate data blocks for the following situations:

\begin{figure}
  \begin{lstlisting} 
   aee8:   cmp r0, #211
   aeec:   bhi 0xd170
   aef0:   adr r1, 0xaef8
   aef4:   ldr pc, [r1, r0, LSL 2] 
   aef8:   .word 0xcbc8
   ...
   b16c:   .word 0xb248
  \end{lstlisting}
  \caption{Jump table snippet in \ArmISA{} binutils's \texttt{strip} binary.  The
    instruction at \lstinline{0xaef0} loads the jump table start address 
     (\lstinline!0xaef8!), \lstinline!r0! corresponds to
    the index variable, and the shift \lstinline{LSL 2} implies that
    each jump table entry is $4$ bytes.  The comparison at address
    \lstinline{aee8} indicates that the jump table has $211$ entries.
  }
  \label{asm:jump-table-example}
  \end{figure}


\begin{figure*}

\begin{tikzpicture}
    \tikzset{
        addr/.style={
          rectangle, draw, minimum height=1cm, minimum width=3cm, fill=blue!10, text centered
        }
      }
      \node[anchor=south west] at(0,8em) {Address};
      \draw[gray] (0, 8 em+2pt) -- (17.5, 8 em+2pt);
      \foreach \y/\label in {
        7/19754:, 
        6/19756:, 
        5/19758:, 
        4/1975a:, 
        3/1975c:, 
        2/1975e:, 
        1/19760:, 
        0/19762: } {
            \draw[dotted] (0, \y em +2pt) -- (17.5, \y em+2pt);
            \node[anchor=south west] at (0, \y em) {\label};
        }

    \node[anchor=south west] at(1.5,8em) {Data};
    \foreach \y/\label in {
        7/bcd6, 
        6/0200, 
        5/1340, 
        4/2de9, 
        3/6040, 
        2/9fe5, 
        1/0400, 
        0/8de5 } {
            \node[anchor=south west] at (1.5, \y em) {\label};
        }

        \node[anchor=south west] at(2.5,8em) {ARM};
        \foreach \y/\asm in {
            7/\lstinline{strheq sp,[r2],-ip}, 
            5/\lstinline{push	\{r0,r1,r4,lr\}},
            3/\lstinline{ldr	r4,[pc,\#96]}, 
            1/\lstinline{str	r0,[sp,\#4]}
            } {
                \node[anchor=south west] at (2.5, \y em) {\asm};
            }

            \node[anchor=south west] at(5.8,8em) {Thumb};
    \foreach \y/\label in {
        6/\lstinline{movs r2,r0}, 
        5/\lstinline{ands   r3,r2}, 
        4/\lstinline{push.w \{r5,r6,lr\}}, 
        2/\lstinline{b      \#0x192a0}, 
        1/\lstinline{movs   r4,r0},
        0/\lstinline{b      \#0x19280}
        } {
            \node[anchor=south west] at (5.8, \y em) {\label};
        }

        \newcommand{\bpos}{10}
        \node[anchor=south west] at(\bpos,8em) {Candidate blocks};

        \node[fill=gray!10,draw,rectangle,minimum width=1em,anchor=south west,text height=1.5em,inner sep=0,text depth=0.5em-1pt] at(\bpos,6em+2pt) {D};
        \node[fill=white,draw,rectangle,minimum width=1em,anchor=south west,text height=1.5em,inner sep=0,text depth=0.5em-1pt] at(\bpos+0.5,6em+2pt) {A};
        \node[fill=gray!10,draw,rectangle,minimum width=1em,anchor=south west,text height=3.5em,inner sep=0,text depth=2.5em-1pt] at(\bpos+1,0em+2pt) {A};
        \node[fill=white,draw,rectangle,minimum width=1em,anchor=south west,text height=0.8em,inner sep=0,text depth=0.2em-1pt] at(\bpos+1.5,6em+2pt) {T};
        \node[fill=white,draw,rectangle,minimum width=1em,anchor=south west,text height=0.8em,inner sep=0,text depth=0.2em-1pt] at(\bpos+2,5em+2pt) {T};
        \node[fill=white,draw,rectangle,minimum width=1em,anchor=south west,text height=2em,inner sep=0,text depth=1em-1pt] at(\bpos+1.5,2em+2pt) {T};
        \node[fill=white,draw,rectangle,minimum width=1em,anchor=south west,text height=0.8em,inner sep=0,text depth=0.2em-1pt] at(\bpos+2,1em+2pt) {T};
        \node[fill=white,draw,rectangle,minimum width=1em,anchor=south west,text height=0.8em,inner sep=0,text depth=0.2em-1pt] at(\bpos+1.5,0em+2pt) {T};

        \newcommand{\spos}{14}
        \node[anchor=south west] at(\spos,8em) {Sorted by end address};

        \node[fill=gray!10,draw,rectangle,minimum width=1em,anchor=south west,text height=1.5em,inner sep=0,text depth=0.5em-1pt] at(\spos,6em+2pt) {2};
        \node[fill=white,draw,rectangle,minimum width=1em,anchor=south west,text height=1.5em,inner sep=0,text depth=0.5em-1pt] at(\spos+0.5,6em+2pt) {1};
        \node[fill=gray!10,draw,rectangle,minimum width=1em,anchor=south west,text height=3.5em,inner sep=0,text depth=2.5em-1pt] at(\spos+1,0em+2pt) {7};
        \node[fill=white,draw,rectangle,minimum width=1em,anchor=south west,text height=0.8em,inner sep=0,text depth=0.2em-1pt] at(\spos+1.5,6em+2pt) {3};
        \node[fill=white,draw,rectangle,minimum width=1em,anchor=south west,text height=0.8em,inner sep=0,text depth=0.2em-1pt] at(\spos+2,5em+2pt) {4};
        \node[fill=white,draw,rectangle,minimum width=1em,anchor=south west,text height=2em,inner sep=0,text depth=1em-1pt] at(\spos+1.5,2em+2pt) {5};
        \node[fill=white,draw,rectangle,minimum width=1em,anchor=south west,text height=0.8em,inner sep=0,text depth=0.2em-1pt] at(\spos+2,1em+2pt) {6};
        \node[fill=white,draw,rectangle,minimum width=1em,anchor=south west,text height=0.8em,inner sep=0,text depth=0.2em-1pt] at(\spos+1.5,0em+2pt) {8};


\end{tikzpicture}
\caption{\ArmISA{} Binary snippet extracted from program \texttt{procd} from the openwrt dataset~\cite{arm-evaluation}.
The example illustrates different possible interpretations as Data (D), ARM (A), or Thumb (T) code of the address range
$[19754,19764)$.
The right hand side contains a representation of all the candidate blocks generated by the extended traversal,
annotated with the decode mode (left) and numbered according to their end address (right).
The real blocks are gray.
}
\label{fig:running-example}
\end{figure*}
%
%
%
%
%
%
%
%
%
%
%
%

\begin{itemize}
\item \textbf{Jump tables} We have multiple rules to detect different
  kinds of jump tables. Most jump tables follow a three-step pattern:
  (1) load the jump table start address, (2) use an index
  variable to compute the address of a jump table entry and load its
  content, and (3) perform an indirect jump to the computed address.

  Our detection rules focus on extracting the jump table start address,
  the size of the jump table entries, and the number of entries in the
  jump table.

  \begin{example}
 Figure~\ref{asm:jump-table-example} contains an example of a jump
 table in \ArmISA{} in which the starting point, entry size, and number of
 entries can be identified.  Our algorithm will generate a data block
 candidate spanning from \lstinline!0xaef8! to
 \lstinline!0xaef8!$+(212*4)=$\lstinline!0xb248!.
 \end{example}
  
  Unfortunately, it  is sometimes challenging to determine a jump table size (the
  number of entries). If the jump table size cannot be determined,
  we start generating data block
  candidates at the jump table start address and with the size of the jump
  table entry, and continue generating candidates sequentially as long
  as:
   (1) The corresponding jump table entry points\footnote{Note that a
    jump table entry does not necessarily contain an absolute
    address. It might contain a relative address in which case the
    target needs to be computed accordingly.} to potentially valid
      code.
   (2) The jump table entry does not overlap with any of the already traversed
      jump table targets or with another jump table start.
      
  This approach might overestimate the length of
  jump tables and result in spurious data block
  candidates which can be resolved by the conflict
  resolution phase. In practice, we believe this happens rarely since jump
  table targets are often located right after the jump table.

  \begin{example}
    Let us consider the jump table in
    Figure~\ref{asm:jump-table-example} again.  If the jump table size
    was not inferred from the comparison at address
    \lstinline{0xaee8}, our algorithm will generate data block
    candidates of size 4 starting at \lstinline{0xaef8} and at 4 byte
    increments. At address \lstinline{0xaef8}, we know the target of
    the jump table entry \lstinline{0xcbc8} points to code, and thus
    must be an upper bound on the extent of the jump table.  As we
    traverse subsequent jump table entries, we tighten that upper
    bound with the discovered jump table targets. Once we reach
    the entry at address \lstinline{b16c}, we update the jump table
    limit to \lstinline{0xb248} which is indeed the end of the jump
    table.
    
  \end{example}
  
\item\textbf{Potential strings} Whenever we detect a memory access to
  a location that could contain a string (a sequence of valid ASCII
  characters that ends with $\backslash$0), we create a data block
  candidate with the size of the potential string.  For potential
  strings above a threshold ($>$8 bytes), we generate data block
  candidates even if no reference from the code is detected.
\item \textbf{Repeated bytes} Whenever we find a sequence of repeated
  bytes over a certain length ($>$8 bytes), we create a data block
  candidate encompassing all the repeated bytes.
\item \textbf{Other data accesses} Data accesses from the code that do not
  correspond to jump tables or strings fall in this category. In such
  cases, we create a candidate data block using the size of the data
  access.
Note that it is often necessary to consider multiple instructions to
determine memory accesses, especially for
RISC architectures like \ArmISA{}. 

%
%

\end{itemize}

\paragraph{\textbf{Extended linear traversal}}
The linear traversal continues traversing code after the end of code blocks.
Our extension accounts for \ArmISA{} decode modes and data blocks.

The linear traversal in \ArmISA{} maintains the decode mode, i.e. it will
attempt to decode instructions using the same decode mode ($A$ or
$T$) as the last visited code block.  However, if the traversal
encounters invalid instructions (as determined by the backward
traversal), it will try switching decode mode (e.g., from ARM to Thumb or vice-versa).

A linear traversal can cross padding instructions but might be
interrupted by data interleavings (since data interleaving might not
be interpretable as instructions). Thus, we initiate linear traversal
\emph{after all} candidate data blocks as well. For \ArmISA{}, traversal after
candidate data blocks is attempted with both ARM and Thumb decode
modes.

\begin{example}
The example in Figure~\ref{fig:running-example} contains a candidate data block
$\DataBlock(19754,19758)$. This data block will trigger extended code traversals 
starting at address $19758$ in both ARM and Thumb mode, resulting in candidate
blocks 7, 4, 5, 6, and 8, respectively (see numbering on the right-hand side of Figure~\ref{fig:running-example}).

\label{example:traversal}
\end{example}

\subsubsection{Sound Overapproximation}
\label{sec:candidate-errors}

In this section, we formally define the conditions that make
a candidate block set a sound overapproximation, discuss its failure modes, and how
our candidate generation phase can result in errors.

\begin{definition}[Sound Candidate Overapproximation]
  \label{def:sound-candidate-overapproximation}
  A candidate block set $\CandidateBlocks$ is a \emph{sound overapproximation} of the code
  if there exists a subset $\BlockSet \subset \CandidateBlocks$ such that
  $\bigcup_{b\in \BlockSet} \BlockInsn(b) = \Code$. 
\end{definition}

There are two situations that can result in a  $\CandidateBlocks$
set that is not a sound overapproximation.

\paragraph{\textbf{Missed instructions}}
  There is some $\instr(s,d)\in \Code$
  that does not belong to any candidate code block.  This can happen
  if the forward code traversal does not visit those instructions.
  Missed instructions will inevitably lead to false negatives, i.e. real
instructions that are not considered as code.

\paragraph{\textbf{Incorrect code block boundaries}} There is a candidate code block $b$
  in which some instructions are code and others are not, i.e.
  $\BlockInsn(b) \cap \Code \neq \emptyset$ and $\BlockInsn(b)\not\subseteq \Code$. 
  This situation can happen if a sequence of spurious instructions falls through a
  sequence of real instructions, and no block limit can be inferred at the boundary.
  Incorrect code block
  boundaries will lead to false positives if the candidate block with
  incorrect boundaries is selected during conflict resolution or false
  negatives otherwise. 

\begin{example}
Consider the example in Figure~\ref{fig:running-example}. If instruction $\instr(19754,T)$ was code,
 it would be a missed instruction since it does not belong to any candidate code block.

If no traversal started at address $19758$ (as described in Example~\ref{example:traversal}), ARM instruction at $19754$ would fall through
to $19758$, generating a single candidate block $\CodeBlock(19754,19764,A)$ instead of the two
separate candidates 2 and 7. Such a block would have incorrect boundaries since it would contain
both real (at $[19758-19764)$) and spurious instructions (at $19754$).
\end{example}

\begin{theorem}
If a candidate block set $\CandidateBlocks$ does not present missed instructions nor incorrect
code block boundaries, then $\CandidateBlocks$ is a sound overapproximation.
\label{theorem:sound-overapproximation}
\end{theorem}

Proofs for all theorems can be found in Appendix~\ref{app:proofs}.

The traversal outlined in this section does not guarantee that we generate a sound
overapproximation. However, this characterization allows us to distinguish disassembly failures caused by the 
candidate generation phase as opposed to failures due to incomplete heuristics or inadequate
heuristic weights. We measure the prevalence of these errors (missed instruction and incorrect block boundaries) experimentally in
Sect.~\ref{sec:experiments:candidates}.

\subsection{Block Weight Assignment}
\label{sec:analysis}

The goal of this phase is to compute the weight associated with each candidate
block ($\bweight(b)$ for each $b\in \CandidateBlocks$).
  
Our analysis defines a set of heuristic rules $h_j
\in \heuristicSet$ that can match individual candidate blocks one or several times
$\heuristicMatch: \CandidateBlocks\times H \rightarrow \mathbb{N}$ (where $\mathbb{N}$ is the set of natural numbers).
Each heuristic $h_j$ has a corresponding integer weight $w_j \in \mathbb{Z}$.
We distinguish two kinds of rules depending on how they contribute to a block's weight: simple and proportional. Simple heuristics ($\heuristicSets$) 
contribute weight based solely on the number of matches and the heuristic weight, 
whereas proportional heuristics ($\heuristicSetp$) contribute proportionally to the size of the candidate block in bytes.

Let $b\in \CandidateBlocks$, its overall weight is computed as follows.
\begin{equation}
  \bweight(b) = \sum_{h_j\in H_s}\heuristicMatch(b,h_j) w_j +  \sum_{h_k\in H_p}\heuristicMatch(b,h_k)   \blockSize(b)w_k
  \label{def:weight}
\end{equation}

We extended Ddisasm to have $34$ multi-ISA heuristics and $58$ \ArmISA{}-specific heuristics.
While enumerating all the heuristics falls beyond the scope of this paper, heuristics can be grouped into the following categories:
\begin{itemize}
\item \emph{Control flow}: Rules that assign points based on how a candidate code block is referenced or references other candidate code blocks.
\item \emph{Instruction patterns}: Rules that assign points based on likely or unlikely instruction patterns.
\item \emph{Metadata}: Rules that use existing metadata (whenever available) such as symbols, relocations, or exception information. 
For example, instructions overlapping with function symbols receive negative points.
  
\item \emph{Data references}: Rules that assign points to candidate data blocks that are referenced
  by candidate code blocks. This includes rules that correspond to jump table detection and literal pool accesses.
  
\item \emph{Data content}: Rules that assign points to candidate data blocks based on their content.
\item \emph{Surrounding context}: Rules that assign points to
  candidate blocks based on the surrounding candidate blocks. E.g. a
  code block that fits in between two other candidate code blocks is
  more likely to be real, or a literal pool entry is often surrounded
  by other literal pool entries.
\end{itemize}


\subsection{Conflict Resolution}

 The conflict resolution phase selects a
subset of blocks from $\BlockSet\subseteq\CandidateBlocks$
that maximizes the overall weight and such that there are no
overlaps among the selected blocks (relying on the assumptions
described in Sect.~\ref{sec:assumptions}).
Prefix-enclosed instructions (Example~\ref{example:prefix-enclosed})
are handled as a special case \footnote{Prefix-enclosed instructions
result in two overlapping candidate blocks in which their first instruction has a prefix in one and no prefix in the other.
We consider only one of the blocks (the one with a prefix) for conflict resolution. Whichever decision is taken for that candidate block
(whether it is included in the final $\BlockSet$) is also adopted for the other candidate in a post-processing phase.}.
For the remainder of the paper, we assume that selected blocks $\BlockSet$ do not overlap with each other.

Maximizing the weights of the selected blocks while avoiding overlaps is a problem
that can be encoded directly as a weighted interval scheduling problem.
In previous sections, we have described how to compute a set of
candidate blocks $b \in \CandidateBlocks$ and assign weights to each candidate block $\bweight(b)$.
These correspond to the input of
the algorithm.

\begin{algorithm}
\caption{Block Conflict Resolution}\label{alg:weightedinterval}
\begin{algorithmic}[1]

  \Function{IntervalScheduling}{$\CandidateBlocks$}
  \State $b_1, b_2, \ldots, b_n$ = \Call{SortByEnd}{$\CandidateBlocks$}
  \Function{Pred}{$i$}
    \State \Return $\max(\{0\}\cup \{ j ~|~ end(b_j) \leq start(b_i)\}) $
  \EndFunction
    \State // Compute maximum weight
    \State $\opt[0] = 0$ 
    \For{$i = 1$ to $n$}
    \State $\opt[i] = \max(\bweight(b_i) + \opt[\Call{Pred}{i}], \opt[i-1])$
            \If{$\bweight(b_i) + \opt[\Call{Pred}{i}] \geq \opt[i-1]$} \label{line:tie}
            \State $\mathit{mem}[i] = \Call{Pred}{i}$
        \Else
            \State $\mathit{mem}[i] = i-1$
        \EndIf
        \EndFor

    \State // Recover optimal schedule
    \State $\BlockSet=\emptyset$ 
    \State $i = n$
    \While{$i > 0$}
    \If{$\mathit{mem}[i] == \Call{Pred}{i}$ and $\bweight(b_i)\geq 0$ }
      \State  $\BlockSet= \BlockSet\cup \{b_i\} $
      \State $i = \mathit{mem}[i]$
    \Else
      \State $i = i-1$
    \EndIf
    \EndWhile 
\State \Return{$\BlockSet$}
\EndFunction

\end{algorithmic}
\end{algorithm}

Each candidate block $b := block([s,e),d)$ corresponds to a task that
needs to be ``scheduled" which starts at address $\instrStart(b) = s$ and ends
at address $\instrEnd(b)= e$.
The output of the algorithm is a set of selected blocks  $\BlockSet$ that
form the optimal schedule.

Algorithm~\ref{alg:weightedinterval}'s implementation closely follows the textbook dynamic
programming implementation of a weighted scheduling algorithm
(e.g. \cite{kleinberg}).
The algorithm contains three high-level steps.  First, we sort
candidate blocks by increasing \textit{end address}. If we have several
candidate blocks with the same \textit{end address} we consider (1) the start
address and (2) the block mode ($A<T<D$) lexicographically to ensure a total order.
Figure~\ref{fig:running-example} presents an example of this ordering on its right-hand side.
Given this ordering, we can compute the \emph{predecessor} index $\Call{Pred}{i}$ for each
block $b_i$. $b_i$'s predecessor index refers to the last block that
precedes $b_i$ and does not overlap with it (given the ordering of
blocks by end address, we know that all preceding blocks do not overlap either).

\begin{example}
  $\Call{Pred}{i}$ has the following values for the candidate blocks in Figure~\ref{fig:running-example}:

  \noindent\begin{tabular}{c@{\,}@{\,}c@{\,}@{\,}c@{\,}@{\,}c}
    $\Call{Pred}{1}=0$ & 
    $\Call{Pred}{2}=0$ & 
    $\Call{Pred}{3}=0$ & 
    $\Call{Pred}{4}=3$ \\
    $\Call{Pred}{5}=4$ &
    $\Call{Pred}{6}=5$ & 
    $\Call{Pred}{7}=3$ &
    $\Call{Pred}{8}=6$ \\
  \end{tabular}
\end{example}
Second, the dynamic programming algorithm computes the maximum
weight for all the scheduling subproblems up to block $b_i$ in $opt[i]$. Each
iteration considers whether a candidate block is included in the
schedule or not. At each step, $mem[i]$ records that 
choice\footnote{In case of a tie, 
Algorithm~\ref{alg:weightedinterval} selects the
 latter block (see $\geq$ in Line~\ref{line:tie}).}.
For example, in the interval scheduling problem from Figure~\ref{fig:running-example},
we will have $\bweight(b_7)+ opt[3] > \opt[6]$, which results in $b_7$ being
chosen over the optimal schedule up to block 6. Thus, we have
$\opt[7]=\bweight(b_7)+ opt[3]$ and $\mathit{mem}[7]=3$.
Finally, in the third step, we recover the optimal schedule
(the final block set) by
traversing $\mathit{mem}[i]$ backwards from the last candidate block.
  When $\Call{Pred}{i} = i-1$, 
  the choice is between including $b_i$ in the schedule or not, which will depend on
  whether $b_i$'s weight is positive or negative.


\section{Heuristic Weight Assignment}
\label{sec:weight_assignment}

Our disassembly algorithm (Sect.~\ref{sec:algorithm}) assumes each
heuristic $h_j\in H$ has a corresponding integer weight $w_j \in
\mathbb{Z}$. In its initial implementation those weights were
hand-picked based on the programmer's intuition and on manual
evaluations. Unfortunately, this approach does not scale, and weight
assignment becomes harder as new heuristics are developed
or updated. As mentioned in the introduction, this problem is not
exclusive to Ddisasm, but common to approaches that rely on weighted
heuristics such as D-ARM~\cite{d-arm}.

In this section, we describe our automated approach for assigning
optimal weights to the disassembly heuristics to maximize the
disassembly's accuracy.
This approach relies on having a collection of binaries annotated with
ground truth information. In particular, it assumes the existence of
\emph{partial} ground truth information consisting of a set of
instructions $\TrueCode$ and a set of
addresses $\Ignored$ for which ground truth
  is not available. 
We discuss how ground truth is extracted for each dataset in Sect.~\ref{sec:experiments},
and the specific limitations that result in $\Ignored$ addresses in Appendix~\ref{app:gt-limitations}.

\subsection{Overview}
\label{sec:weights-overview}
Given a binary with ground truth, we can partially run the
disassembly algorithm. This allows us to collect candidate blocks
$\CandidateBlocks$ and heuristic matches
$\heuristicMatch(b,h)$ for each candidate block $b\in
\CandidateBlocks$ and each heuristic $h \in H$.

In addition, we can map our ground truth information $\TrueCode$
and $\Ignored$ to two subsets of candidate blocks:
$\TrueBlocks \subset \CandidateBlocks$ and  $\FalseBlocks \subset \CandidateBlocks$ (``True'' and ``False''
blocks respectively).
We define $\TrueBlocks$ as:
$$
  \TrueBlocks = \bigcup_{i\in \TrueCode} \BlockOf(i)
$$

Let $\instrStart(\BlockInsn(b))$ denote the set of all starting addresses
  of instructions in a code block $b$. We then define $\FalseBlocks$ as follows:
\begin{multline*}
  \FalseBlocks = \{b ~|~  b\in \CandidateBlocks \setminus \TrueBlocks \\
       \land \instrStart(\BlockInsn(b)) \not\subseteq \Ignored\}
\end{multline*}

If $\CandidateBlocks$ is a sound overapproximation (Definition~\ref{def:sound-candidate-overapproximation}),
$\TrueBlocks$ contains all and only true instructions ($\TrueCode$).
$\FalseBlocks$ are candidate code blocks that are neither true blocks
nor completely ignored. Note that there are candidate blocks
that are neither $\TrueBlocks$ nor $\FalseBlocks$. Those are blocks for which we do not have
definitive ground truth.
In particular, this applies to all the candidate data blocks\footnote{As opposed to candidate code blocks, which are guaranteed not to share instructions 
(Sect.~\ref{sec:original-traversal}), we do not enforce any properties on candidate data blocks. For
a region of the binary that should be considered data, there might be multiple competing data blocks and our ground truth
does not tell us which ones to pick.}.

\begin{example}
In our example in Figure~\ref{fig:running-example}, $\Code=\{\instr(19758,A), \instr(1975c,A), \instr(19760,A)\}$
and $\Ignored = \emptyset$, 
which results in $\TrueBlocks=\{b_7\}$ and $\FalseBlocks=\{b_1,b_{3-6},b_8\}$.
Candidate data block $b_2$ does not belong to either.
\label{example:weight-inference-sets}
\end{example}

Rather than performing interval scheduling (conflict resolution) to
find the block selection based on a given set of fixed weights, we need to
find out which weights will lead to the right decisions during
interval scheduling. The correct decisions involve selecting all
candidate blocks in $\TrueBlocks$ and not selecting any
blocks in $\FalseBlocks$. Schedules that satisfy
those conditions are \emph{optimal}.

\begin{definition}[Optimal schedule]
  A candidate block set $S$ is an \emph{optimal} schedule if $\TrueBlocks \subseteq S$ and $\FalseBlocks\cap S=\emptyset$.
  \label{def:optimal-schedule}
  \end{definition}

We propose a new algorithm (Sect.~\ref{sec:wis-constraint-inference}) that
performs weighted interval scheduling symbolically (with symbolic
weights) and uses the ground truth information to collect a set of linear
constraints that enforce the selection of an optimal schedule.
For clarity, we first present a naive implementation for
 inferring constraints from a weighted interval
scheduling problem in
Sect.~\ref{sec:wis-constraint-inference}
and later present an optimized implementation in Sect.~\ref{sec:simplification}.

We apply this algorithm to a collection of binaries and infer a
constraint set for each binary. The overall constraint set is not
guaranteed to be satisfiable---It is possible that there is no weight
assignment that satisfies all the constraints, which is an indication
that additional heuristics might be needed or some heuristic rules
might need to be refined.
Thus, we encode the problem as a linear programming (LP) optimization with soft constraints
to find a weight assignment that maximizes the number
of satisfied constraints (see Sect.~\ref{sec:weight-inference}).

\subsection{Weight Constraints Inference}
\label{sec:wis-constraint-inference}

The naive algorithm for inferring constraints (Algorithm~\ref{alg:wis-constraint-inference-naive}) follows
a similar structure as Algorithm~\ref{alg:weightedinterval}.  
They both
iterate over all candidate blocks sorted by their end address.  While
Algorithm~\ref{alg:weightedinterval} collects an optimal numerical value in each iteration
$opt[i]$, the constraint inference algorithm collects
a set of symbolic schedules in 
$\symopt[i]$.

Let $b$ be a candidate block, its symbolic weight is defined as $\sweight(b)$.
We compute it with the same formula as $\bweight(b)$ (Equation~\ref{def:weight})
but using symbolic values for the heuristic weights $w_j$. Instead of an integer, $\sweight(b)$
returns a linear expression of the form 
$c_1 w_1
+ c_2 w_2 + \ldots + c_m w_m$ where $c_j \in \mathbb{N}$ and $w_1,
w_2, \ldots, w_m$ are variables representing the unknown heuristic
weights. 

Let $S$ be a set of blocks representing a concrete schedule, we represent
in our algorithm with a tuple $\langle\sweight(S),\score(S)\rangle$.
The first term $\sweight(S)$
represents the symbolic weight of the schedule, defined as $\sweight(S)=\sum_{b\in S} \sweight(b)$.
The second term
$\score(S)$ is a numerical score
based on the number of $\TrueBlocks$ and $\FalseBlocks$ that are included in
the schedule.
\begin{equation}
  \score(S) =
  \begin{cases} 
  -1 & \text{if } \FalseBlocks \cap S \neq \emptyset \\
  |\TrueBlocks \cap S| & \text{otherwise}
  \end{cases}
  \label{def:score}
\end{equation}

This score captures our priority in selecting schedules.
Optimal schedules are guaranteed to have maximal score.
\begin{lemma}
Let $S$ be a schedule, if $S$ is optimal (Definition~\ref{def:optimal-schedule}), 
its score is $\score(S)=|\TrueBlocks|$, otherwise $\score(S)<|\TrueBlocks|$.
\label{lemma:optimal-score}
\end{lemma}
Lemma~\ref{lemma:optimal-score} follows directly from the definition of $\score$ (Equation~\ref{def:score}).


\begin{algorithm}
  \caption{Naive Constraint Inference}\label{alg:wis-constraint-inference-naive}
  \begin{algorithmic}[1]
    \Function{InferCS}{$\CandidateBlocks,\TrueBlocks$,$\FalseBlocks$}
      \State $\symopt[0] = \{\langle 0,0\rangle\}$
      \For{$i = 1$ to $n$}
      \State $\mathit{take}$= $\{ \langle\sweight(b_i) +s.\SSchedWeight,\updateScore(s.\SSchedScore,b_i)\rangle$\\
      $\quad\quad\quad\quad\quad ~|~ s \in  \symopt[\Call{Pred}{i}] \}$
      \State $\mathit{leave}$ = $\symopt[i-1]$
          \State $\symopt[i] = \mathit{take} \cup \mathit{leave}$
        \EndFor
        \State $l$ = $\bestSchedule(\symopt[n])$
        \State $rs$ = $\{ r ~|~ r \in \symopt[n] \land r.\SSchedScore < l.\SSchedScore \}$ $\label{line:non-optimal}$
        \State \Return $\{ l.weight > r.weight ~|~ r \in rs\}$ $\label{line:cs}$
    \EndFunction
  \end{algorithmic}
  \end{algorithm}

Algorithm~\ref{alg:wis-constraint-inference-naive}'s main loop amounts to
the incremental computation of all the symbolic schedules.
Let $s$ be a symbolic schedule, we use $s.\SSchedWeight$ and $s.\SSchedScore$ to refer to the weight and score of $s$ respectively.
On each iteration $i$, we have a decision point: whether $b_i$ is
included in the schedule.  We compute the two alternatives
$\mathit{take}$ (when $b_i$ is included) and $\mathit{leave}$ (when $b_i$ is not
included). To compute $\mathit{take}$ we need to add a symbolic expression
corresponding to block  $b_i$'s weight to all the schedules in
$\symopt[\textsc{Pred}(i)]$ (we reuse \textsc{Pred}'s definition from Algorithm~\ref{alg:weightedinterval}). 

The score of the schedules in $\mathit{take}$ is updated using
$\updateScore$ which corresponds to the incremental computation of $\score$ (see Equation~\ref{def:score}).
Let $s$ be a score and $b$ a block under consideration, $\updateScore$ is defined as follows:
 \begin{equation}
\updateScore(s,b) =
\begin{cases} 
-1 & \text{if } s=-1 \lor b\in \FalseBlocks\\
s+1 &  \text{if } b\in \TrueBlocks\\
s & otherwise
\end{cases}
\label{def:udate_score}
\end{equation}

 The value of $\symopt[i]$ corresponds to the maximum of all the
 potential schedules in $\mathit{take}$ and $\mathit{leave}$.  
 Once we have computed $\symopt$, $\symopt[n]$ contains a symbolic
 representation of all the possible schedules (block selections) for
 the binary.  In particular, it contains at least one optimal schedule.
 Given that our ground truth is partial 
 and there can be overlapping candidate data blocks,
 there can be several optimal schedules. 
 Function $\bestSchedule$ returns an optimal schedule by leveraging Lemma~\ref{lemma:optimal-score}.
 If there are several, it
 chooses one of them heuristically by selecting the optimal schedule with the highest sum of positive heuristic
 coefficients.
 Then, we generate linear constraints that ensure that our selected optimal schedule
 has a higher overall weight than every other non-optimal schedule
 (which are guaranteed to have a lower score by Lemma~\ref{lemma:optimal-score}).
 
\begin{example}
  \label{example:constraints-naive}
Continuing from Example~\ref{example:weight-inference-sets}, 
let us assume we have 4 heuristic rules $s$ (size), $j$ (jumped), $l$ (literal pool), and $c$ (called)
with unknown weights $w_s$, $w_j$, $w_l$, and $w_c$ respectively.
Assume further that $s$ is a proportional rule and all the others are simple,
each block matches $\heuristicMatch(b,s)=1$, and we have $\heuristicMatch(b_2,l)=1$,
  $\heuristicMatch(b_3,j) = 1$, $\heuristicMatch(b_5,j)=1$, and $\heuristicMatch(b_7,c)=1$. All other combinations
  of blocks and rules yield zero.
In this scenario, we have the following symbolic weights for each of the blocks:

\noindent\begin{tabular}{@{}l@{\quad\quad}l}
  $\sweight(b_1) = 4w_s$ & $\sweight(b_5) =  6w_s+ w_j$ \\
  $\sweight(b_2) = 4w_s+ w_l$ & $\sweight(b_6) =2w_s$   \\
  $\sweight(b_3) = 2w_s+w_j$ & $\sweight(b_7) = 12w_s+ w_c$  \\
  $\sweight(b_4) = 2w_s$  & $\sweight(b_8) =2w_s $ 
\end{tabular}

  At the end of the loop, $\symopt[8]$ contains symbolic
  schedules representing all the combinations of blocks.
  In particular, it contains a tuple representing
  the optimal schedule $\{b_2,b_7\}$:
    $\langle16w_s+w_l+w_c,1\rangle$, as well as non-optimal
    schedules,  such as $\{b_3,b_7\}$: $\langle14w_s+w_j,-1\rangle$,
    which should not be selected.
    Based on those two schedules, Algorithm~\ref{alg:wis-constraint-inference-naive} generates
   a linear constraint $16w_s+w_l+w_c > 14w_s+w_j$.
   Note that even in this example, there are two optimal schedules $\{b_7\}$  and $\{b_2,b_7\}$.
   
\end{example}

\begin{theorem}
  Let $\constraintSet$ be the set of constraints generated by Algorithm~\ref{alg:wis-constraint-inference-naive} and
  let $\alpha: H \rightarrow \mathbb{Z}$  be a weight assignment that satisfies the constraints $\alpha \models \constraintSet$, 
then there is an optimal schedule $S$  such that $\alpha(\sweight(S))> \alpha(\sweight(S'))$ for every non-optimal schedule $S'$.
\label{theorem:infer-constraint-correctness}
\end{theorem}

\begin{collorary}
  Let $\constraintSet$ be the set of constraints generated by Algorithm~\ref{alg:wis-constraint-inference-naive} and
let $\alpha: H \rightarrow \mathbb{Z}$  be a weight assignment that satisfies the constraints $\alpha \models \constraintSet$,
  then the schedule selected by Algorithm~\ref{alg:weightedinterval} using 
 weight assignment $\alpha$ is optimal.
\end{collorary}

This theorem and corresponding corollary ensure that a satisfying weight assignment
will lead to an optimal schedule---an optimal block selection with respect to the ground truth---being
selected during conflict resolution.

The fact that we choose only one optimal schedule in $\bestSchedule$ 
and that optimal schedules are not necessarily unique (see Example~\ref{example:constraints-naive})
means that our algorithm is not complete. Even if we fail to
find a satisfying weight assignment that selects an optimal schedule,
there might exist one for a \emph{different} optimal schedule.
This is acceptable for our use case since our goal is to learn heuristic weights
that work well in practice and minimize the number of errors even in situations where errors cannot
be completely eliminated.

\subsection{Optimized Constraint Inference}
\label{sec:simplification}

The number of potential schedules in $\symopt[i]$ can, in the worst case, double in each
 iteration. We alleviate this exponential growth risk by incorporating
several optimizations. Algorithm~\ref{alg:wis-constraint-inference-opt}
contains the most relevant optimizations.

\subsubsection{Interval Scheduling Decomposition}
\label{sec:scheduling-decomposition}
Up to this point, we have considered the block conflict resolution as a single
WIS problem that encompasses all candidate blocks in a binary. However,
WIS problems can often be decomposed into smaller subproblems.

\begin{example}
  Consider the interval scheduling problem in Figure~\ref{fig:running-example}.
  There are many possible schedules, e.g. 
  $\{1,7\}$, 
  $\{2,7\}$, 
  $\{1,4\}$, or
  $\{2,4\}$.
  However, this problem can be decomposed into two completely independent decisions: (i)
  which intervals are chosen between from 1-3 (ii) which ones are chosen from 4-8.
  None of the choices for blocks in 4-8 preclude us from choosing any of the
  blocks from 1-3.
  This means that we split this scheduling problem into two and generate constraints
  for each of the subproblems.
\end{example}

More formally, we can split an interval scheduling problem $b_1,b_2,\ldots b_n$
on $b_i$ if all $\Call{Pred}{j}>= i$ for $i< j \leq n$. In that case, we can recursively unfold
$\symopt[n]$'s definition until every term contains $\symopt[i]$, which can be factored out.
Thus, we
can compute the optimal schedule for $b_1,b_2, \ldots, b_i$ and for $b_{i+1},b_2, \ldots, b_n$
separately and then combine them. In practice, an interval scheduling problem for a binary
can often be split into many small independent subproblems.
This splitting can lead to important
savings in Algorithm~\ref{alg:wis-constraint-inference-opt}. The size
of $\symopt$ can grow exponentially (in the worst case) with the number
of intervals considered ($2^n$), so reducing the size of $n$ has a very
significant effect.
For example, the \lstinline{procd} binary selected in Figure~\ref{fig:running-example} has $4926$ candidate
blocks, but after decomposition, the largest scheduling subproblem contains only $11$ candidates.

\subsubsection{Schedule Subsumption}
Despite not knowing the optimal values of the heuristic weights beforehand,
it is reasonable to fix their sign. We can divide  heuristics
into positive and negative heuristics $H = H^+ \cup H^-$
($w_j \geq 0$ if $h_j\in H^+$ and $w_j\leq 0$ if $h_j\in H^-$).
This distinction allows us to perform further optimizations.

Let $e = c_1w_1+ c_2w_2+\ldots+c_m w_m$ be a linear expression where
$c_j\in \mathbb{Z}$ and $w_j$ are variables representing weights.
We know that $e$ is positive ($e\geq 0$) 
if  $c_j \geq 0$ for all positive heuristics $h_j\in H^+$ and $c_j \leq 0$
for all negative heuristics $h_j\in H^-$.
Then, given two partial schedules $A$ and $B$, we say $A$ \emph{subsumes} $B$ 
if we can prove $\sweight(A)\geq \sweight(B)$ (we know $A$'s weight will be higher than or equal to $B$'s 
regardless of the chosen weights) and $\score(A) \leq \score(B)$.
The second condition $\score(A) \leq \score(B)$ means there are 
three possibilities:
\begin{enumerate}
  \item Both $A$ and $B$ will be part of optimal schedules. In that case  
  $A$ is a better schedule for generating constraints. Any satisfying weight assignment $\alpha$ for $B$, will
also be satisfying for $A$, but there might be additional possibilities for $A$.
  \item Both $A$ and $B$ will be part of non-optimal schedules. Then, 
   a constraint that ensures the weight of $A$ is smaller
than that of an optimal schedule, will also guarantee that $B$'s weight is smaller.
  \item $A$ is part of a non-optimal schedule and $B$ part of an optimal schedule.
  In this case, any constraints trying to enforce $\sweight(B) > \sweight(A)$ will be trivially unsatisfiable.
\end{enumerate}
In all the cases above, we can discard $B$ (we do not need to take it into account for generating constraints).
At each iteration, function $\simplifyS$ removes subsumed schedules 
from $\symopt[i]$ (Line~\ref{line:simplifyS}).

\begin{example}
  In the previous example, we decomposed the scheduling
  from Figure~\ref{fig:running-example} into two, one for candidates
  $1-3$ and another for candidates $4-8$.
  Let us focus now on the latter and assume the same heuristics and heuristic matches
  as Example~\ref{example:weight-inference-sets}. We also assume
  all heuristics are positive.

  Without subsumption, $\symopt[4]$ contains tuples $\langle 0, 0\rangle$, $\langle 2w_s, -1\rangle$ corresponding
  to schedules $\{\}$ and $\{4\}$ respectively.
  Since $w_s$ is positive, we know that $2w_s \geq 0$ and $\langle 2w_s, -1\rangle$ subsumes $\langle 0, 0\rangle$
  which can be discarded.
  After simplifying, $\symopt[4]$ only contains the tuple $\langle 2w_s, -1\rangle$.

  A similar situation happens for $\symopt[5]$. Without subsumption, $\symopt[5]$
    contains 4 tuples corresponding to schedules  $\{\}$, $\{4\}$, $\{5\}$, and $\{4,5\}$.
    However, they are all subsumed by the tuple corresponding to schedule $\{4,5\}$: $\langle 8w_s+w_j,-1\rangle$.
  Thus, the optimized version contains a single tuple.

\end{example}

\begin{algorithm}
  \caption{Optimized Constraint Inference}\label{alg:wis-constraint-inference-opt}
  \begin{algorithmic}[1]
    \Function{InferCs}{$\CandidateBlocks,\TrueBlocks$,$\FalseBlocks$}
      \State $\symopt[0] = \{\langle0,0\rangle\}$
      \State $\constraintSet[0] = \emptyset$ 
      \For{$i = 1$ to $n$}
      \State $\mathit{take}$ = $\{ \langle\sweight(b_i) +s.\SSchedWeight,\updateScore(s.\SSchedScore,b_i)\rangle$\\
      $\quad\quad\quad\quad\quad ~|~ s \in  \symopt[\Call{Pred}{i}] \}$
      \State $\mathit{leave}$ = $\symopt[i-1]$
          \If{$b_i\in \TrueBlocks$}
          \State  $l$ = $\bestSchedule(\mathit{take})$ \label{line:opt1-start}
          \State $\mathit{newCs}$ = $\{ l.\SSchedWeight>r.\SSchedWeight \,|\, r \in \mathit{leave}\}$
          \State $\symopt[i] =  \{ l \}$ \label{line:opt1-end}
          \State  $\constraintSet[i] =\simplifyC(\constraintSet[\Call{Pred}{i}] \cup \mathit{newCs}) $ \label{line:simplifyC1}
          \Else
          \State $\symopt[i] = \simplifyS(\mathit{take} \cup \mathit{leave})$\label{line:simplifyS}
          \State $\constraintSet[i] = \simplifyC(\constraintSet[\Call{Pred}{i}] \cup \constraintSet[i-1])$\label{line:simplifyC2}
          \EndIf
        \EndFor
        \State $l$ = $\bestSchedule(\symopt[n])$
        \State $rs$ = $\{ r | r \in \symopt[n] \land r.\SSchedScore < l.\SSchedScore \}$
        \State $\mathit{newCs}$ = $\{ l.\SSchedWeight > r.\SSchedWeight ~|~ r \in rs\}$
        \State \Return $\simplifyC(\mathit{newCs} \cup \constraintSet[n])$ \label{line:simplifyC3}
    \EndFunction
  \end{algorithmic}
  \end{algorithm}

\subsubsection{Constraint Subsumption}

Similarly to schedule subsumption, we can detect constraint
subsumption.  Let $c_1 := l_1 > r_1$ and $c_2 := l_2 > r_2$ be two
linear constraints. If $(l_2-r_2) - (l_1-r_1) \geq 0$, then $c_1 \Rightarrow c_2$
 and $c_2$ can be discarded.
This can be checked similarly to schedule subsumption by checking
the coefficients of positive and negative heuristics in the 
linear expression $(l_2-r_2) - (l_1-r_1)$.

  Function $\simplifyC$ (Lines~\ref{line:simplifyC1},
  \ref{line:simplifyC2} and \ref{line:simplifyC3}) 
  simplifies the generated constraint sets by removing subsumed constraints.

\subsubsection{Partial Schedule Constraints}
Whenever we encounter a $b_i\in \TrueBlocks$ in the main loop, we know
that $b_i$ must be part of all optimal schedules, and the decision needs
to happen based on the currently accumulated weight. Thus, we can
generate constraints based on those partial schedules and 
discard the partial schedules that do not select $b_i$ and thus are guaranteed to be non-optimal (see
Lines~\ref{line:opt1-start}-\ref{line:opt1-end} in
Algorithm~\ref{alg:wis-constraint-inference-opt}).
These constraints are accumulated in 
$\constraintSet$ throughout
the loop, and we add them to the final constraint set at the end.
This optimization reduces the growth of $\symopt$, given that on each
iteration that corresponds to a true block $b_i \in \TrueBlocks$, we reset
the size of $\symopt[i]$ to $1$. It can also result in a smaller number of 
overall constraints.

\subsection{Solving Weight Constraints}
\label{sec:weight-inference}

We can repeat the procedure described in the previous section
(Sect.~\ref{sec:wis-constraint-inference}) to obtain a constraint
set for each binary with ground truth. We accumulate those
constraints for a collection of binaries and try to find a weight
assignment that satisfies all constraints. However, this might not be
feasible. If an optimal weight assignment does not exist, this is an
indication that the existing heuristics are insufficient to handle all
possible cases. Nevertheless, we want to find an assignment that
minimizes the errors, i.e. that
maximizes the number of satisfied constraints.

We encode this problem as an LP problem with soft constraints.
For each constraint $c_i \in \constraintSet$ of the form $l_i > r_i$, we introduce a positive slack variable $s_i$
and reformulate the constraint as $l_i +s_i \geq r_i+1$. 
Our modified constraint set is denoted as $\constraintSet'$, the set of all slack variables is $\slackSet$,
and our objective function is the sum of the slack variables.
The resulting linear program is:
\[
  \begin{array}{rl}
    \mathrm{minimize} & \sum\limits_{s\in \slackSet} s\\
    \mathrm{subject~to:} & \constraintSet' \\
    &  w_j\geq 0 ~\mathit{for}~h_j\in H^+\\
    &  w_j\leq 0 ~\mathit{for}~h_j\in H^-\\
    &  s \geq 0~\mathit{for}~s\in \slackSet\\\\
  \end{array}
\]
This linear program can be efficiently solved by off-the-shelf solvers (we use
Pulp~\cite{pulp}). Sect.~\ref{sec:experiments} includes experiments that validate our
weight inference approach. 

\section{Experimental Evaluation}
\label{sec:experiments}

In this section, we evaluate our disassembly algorithm and its corresponding
weight inference algorithm on a large collection of binaries with ground truth.
The evaluation is divided into four parts.

First, in Sect.~\ref{sec:experiments:candidates},
we experimentally validate our candidate generation algorithm (Sect.~\ref{sec:candidate-generation}).
Second, in Sect.~\ref{sec:weight-inference-evaluation} we evaluate
our weight inference algorithm and how inferred weights generalize to unseen binaries.
Third, in Sect.~\ref{sec:comp-orig} we evaluate the overall
disassembly effectiveness (precision and recall) and compare it to other state-of-the-art (SOTA) disassemblers.
Fourth, in Sect.~\ref{sec:runtime-performance} we measure the runtime performance of both
our improved disassembly algorithm, and the heuristic weight inference.
All of our experiments are performed on stripped binaries.

\begin{table}[t]
  \caption{Evaluation datasets. Each dataset includes different ISAs and binary formats.
  Each dataset relies on a different method for extracting ground truth. }
  \label{table:datasets}
\begin{tabular}{llll}
Dataset & ISA & Format & Ground Truth \\
\toprule
SOK~\cite{gt-is-not-easy} &  x86/x64,  & ELF & Compiler \\
                          &  \ArmISA{}, \AarchISA{}           &     & Modification                      \\
Pangine~\cite{pangines-gt} & x86/x64 & ELF & Intermediate \\
 &&& Compilation Artifacts\\
Jiang~\cite{arm-evaluation} & \ArmISA{} & ELF & Marker symbols\\
Assemb~\cite{assemblage} & x86/x64 & PE & PDB files\\
\bottomrule
\end{tabular}
\end{table}

\begin{table*}[t]
  \centering
  \caption{Ddisasm Candidate Block Generation Evaluation.}
  \label{table:candidates}


 \begin{tabular}{l|c|r|r|r|r|r|r|r}
   Dataset & ISA & \# Binaries & Failures (\%) &  Sound (\%) & Unsound (\%)& Insns & Missed Insns (\%)& \shortstack{Incorrect\\Blocks}\\
   \hline
   SOK & x86/x64 & 3,974 & 113 (2.84\%) & 3,854 (96.98\%) & 7 (0.18\%) & 179,630,938 & 22 (0.00001\%) & 0 \\ 
   SOK & \ArmISA{} & 2,561 & 49 (1.91\%) & 2,503 (97.74\%) & 9 (0.35\%) & 89,081,800 & 65 (0.00007\%) & 2 \\ 
   SOK & \AarchISA{} & 1,264 & 0 (0\%) & 1,264 (100\%) & 0 (0\%) & 45,484,081 & 0 (0\%) & 0 \\ 
   Pangine & x86/x64 & 879 & 61 (6.94\%) & 816 (92.83\%) & 2 (0.23\%) & 104,427,721 & 0 (0\%) & 2 \\ 
   Jiang & \ArmISA{} & 1,026 & 0 (0.00\%) & 1,025 (99.90\%) & 1 (0.10\%) & 30,297,775 & 4 (0.00001\%) & 0 \\ 
   Assemb & x86/x64 & 85,296 & 6,119 (7.17\%) & 78,659 (92.22\%) & 518 (0.61\%) & 1,458,731,926 & 5,396 (0.00037\%) & 90 \\ 
   \end{tabular}

 \end{table*}

Our experiments leverage four publicly available 
datasets\footnote{We only include the publicly available portion of Jiang's (excluding SPEC binaries) and Assemblage's PE dataset.}. 
Table~\ref{table:datasets}
describes the Instruction Set Architecture (ISA), binary formats, and ground truth source 
for each of the datasets.
These datasets contain binaries compiled with a variety of compilers, including
GCC, Clang, ICC, MSVC, and compiler optimizations O0-O3, Of, and Os. We refer
the readers to the original publications for details.

Pang et al.~\cite{gt-is-not-easy} provide
an extensive discussion on the limitations of each ground truth source.
While modifying the compiler (as in dataset SOK) provides the highest
quality ground truth, this is not feasible for closed-source compilers
such as MSVC or ICC. The Pangine and Assemblage datasets (shortened as Assemb)
complement Pang's dataset with binaries
compiled with those compilers.

During our experiments, we have identified limitations
in the ground truth of each of the considered datasets.
We provide a detailed account of these limitations in Appendix~\ref{app:gt-limitations}.

We have released the code to run all the experiments\footnote{\url{https://github.com/GrammaTech/ddisasm-wis-evaluation}}.

\subsection{Candidate Block Generation}
\label{sec:experiments:candidates}

To substantiate the effectiveness of our algorithm presented in
Sect.~\ref{sec:candidate-generation}, 
we measure the prevalence of missed instructions and incorrect code block boundaries
for each of the datasets.
For each binary, we run our tool \tool{}\footnote{\tool{}'s improvements have been merged into the official Ddisasm repository \url{https://github.com/GrammaTech/ddisasm}. We run our experiments with commit \texttt{\href{https://github.com/GrammaTech/ddisasm/commit/415a2be730311c07b6808fa1243d2dfe8a7f15a9}{415a2be}}.}
with a 1-hour timeout
and collect information about the generated candidate blocks. 
We then compare that information
with the ground truth to compute \emph{missed instructions} and \emph{incorrect code block boundaries} (see Sect.~\ref{sec:candidate-errors}).
A binary's block candidate set is a sound overapproximation if it has neither, it is unsound otherwise.
Thus, a binary can either be sound, unsound, or count as a failure.
Failures include (1) timeouts (51 binaries), (2) \tool{}
failing to produce an output (3473 binaries),
(3) failures in the ground truth extraction from PDBs (273 binaries),
or (4) our scripts
failing to decode an instruction marked as code by the ground truth (2545 binaries).
The latter can be caused by limitations of our instruction decoder (Capstone 5.0.1~\cite{capstone}), or it might signal an error
in the ground truth.

Table~\ref{table:candidates} reports
the number of binaries (and percentage) in each category, the total number of true instructions (Insns),
the number (and percentage) of missed instructions, and the number of candidate blocks with incorrect
boundaries (Incorrect Blocks).
Our extended traversal (Sect.~\ref{sec:extended-traversal}) generates
a sound overapproximation in the vast majority of cases. Consider SOK's dataset. 6 out of the 7 unsound binaries in SOK's x86/x64
dataset correspond to different versions of \lstinline{libc} which present prefix-enclosed instructions (see Figure~\ref{fig:padding}).
While our tool can correctly handle those patterns, they are not currently handled by our evaluation scripts.
As for SOK's \ArmISA{} binaries, our experiments report $9$ unsound binaries, all of them in the Thumb dataset. These binaries have $65$ missing instructions
where $61$ of those
are concentrated in $3$ different versions of the same program (\texttt{ssh-keyscan}).  We believe these
are reported due to errors in the ground truth. 
The overall ratio of unsound binaries is less than  $0.4\%$ across all datasets, with the exception of Assemblage.
We manually examined a sample of unsound binaries and discovered 
both inaccuracies in the PDB file, particularly with jump tables and
obfuscated code, and some limitations of our tool.  We provide additional
details in Appendix~\ref{app:assemblage-pdb}.

\subsection{Weight Inference Evaluation}
\label{sec:weight-inference-evaluation}

To evaluate our weight inference algorithm, we select a random sample of $1,895$ binaries (approximately $20\%$)
of our datasets SOK, Jiang, and Pangines
as our training set. Our algorithm collects $86,564$ constraints and produces an optimal weight assignment using Pulp~\cite{pulp}. 
The optimal weights satisfy $86,263$ constraints, leaving $301$ unsatisfied.
We evaluate the resulting weights in practice by running
\tool{} with those weights on the complete dataset.

We compare the results against a baseline of \tool{} with manually optimized weights.
Note that \tool{}'s manual weights are the result of more than four years of continuous development
and thus we expect them to be close to optimal.
For each binary, we measure the number of correctly recovered instructions (true positives),
spurious instructions (false positives) and the number of true instructions
that were not recovered (false negatives) and use them to compute the overall precision and recall.
These metrics are computed for each dataset, ISA, training and validation subsets, for both manual and learned weights.
The results can be found in Table~\ref{table:weight-inference}.
In addition, we report the number of correct binaries (binaries with no false positives nor
false negatives) in each category.

  \begin{table*}
    \caption{Weight Inference Evaluation: Number of correct binaries, precision, and recall for both the training and validation
    sets, using both manually optimized weights and learned weights.}
    \label{table:weight-inference}
    \begin{tabular}{@{}l|c|c|r|r|r|r|r|r|r|r@{}}
      \multirow{2}{*}{Dataset} & \multirow{2}{*}{ISA} & \multirow{2}{*}{Method}  & \multicolumn{4}{c|}{Training Set} & \multicolumn{4}{c}{Validations Set} \\ 
       & & & \# Binaries  & Correct (\%)  & Precision & Recall & \# Binaries  & Correct (\%)  & Precision & Recall \\ \hline
       \multirow{2}{*}{SOK} & \multirow{2}{*}{x86/x64} &  Manual& \multirow{2}{*}{738} & 687 (93.09\%) & 100.000\% & 99.995\% 
      & \multirow{2}{*}{3,236} & 2,968 (91.72\%) & 99.991\% & 99.997\% \\ 
        &  &  Learned&  & 694 (94.04\%) & 100.000\% & 99.995\% 
      &  & 2,993 (92.49\%) & 99.991\% & 99.997\% \\ \hline
       \multirow{2}{*}{SOK} & \multirow{2}{*}{\ArmISA{}} &  Manual& \multirow{2}{*}{534} & 199 (37.27\%) & 99.965\% & 99.120\% 
      & \multirow{2}{*}{2,027} & 725 (35.77\%) & 99.954\% & 99.906\% \\ 
        &  &  Learned&  & 244 (45.69\%) & 99.964\% & 99.188\% 
      &  & 893 (44.06\%) & 99.952\% & 99.973\% \\ \hline
       \multirow{2}{*}{SOK} & \multirow{2}{*}{\AarchISA{}} &  Manual& \multirow{2}{*}{253} & 252 (99.60\%) & 100.000\% & 100.000\% 
      & \multirow{2}{*}{1,011} & 1,009 (99.80\%) & 100.000\% & 100.000\% \\ 
        &  &  Learned&  & 252 (99.60\%) & 100.000\% & 100.000\% 
      &  & 1,008 (99.70\%) & 100.000\% & 100.000\% \\ \hline
       \multirow{2}{*}{Pangine} & \multirow{2}{*}{x86/x64} &  Manual& \multirow{2}{*}{171} & 163 (95.32\%) & 100.000\% & 100.000\% 
      & \multirow{2}{*}{708} & 658 (92.94\%) & 100.000\% & 99.999\% \\ 
        &  &  Learned&  & 163 (95.32\%) & 100.000\% & 100.000\% 
      &  & 645 (91.10\%) & 100.000\% & 99.999\% \\ \hline
       \multirow{2}{*}{Jiang} & \multirow{2}{*}{\ArmISA{}} &  Manual& \multirow{2}{*}{199} & 148 (74.37\%) & 99.988\% & 99.995\% 
      & \multirow{2}{*}{827} & 635 (76.78\%) & 99.994\% & 99.995\% \\ 
        &  &  Learned&  & 153 (76.88\%) & 99.988\% & 99.996\% 
      &  & 637 (77.03\%) & 99.994\% & 99.996\% \\ \hline
       \multirow{2}{*}{Assemb} & \multirow{2}{*}{x86/x64} &  Manual& \multicolumn{4}{c|}{}
      & \multirow{2}{*}{85,296} & 77,186 (90.49\%) & 99.960\% & 99.996\% \\ 
        &  &  Learned& \multicolumn{4}{c|}{}
      &  & 76,613 (89.82\%) & 99.960\% & 99.994\% \\ \hline
      \end{tabular}

  \end{table*}

  Overall, learned weights achieve similar results as the manually tuned weights in most categories.
  We see improvements in the handling of SOK \ArmISA{} binaries where the recall increases
  from $99.906\%$ to $99.973\%$ (in the validation set),
  which also has an important effect on the number of correct binaries (from $35.77\%$ to $44.06\%$).

  Table~\ref{table:weight-inference} also demonstrates how weights generalize
  to unseen binaries. Despite training in only 20\% of the binaries, all metrics are comparable on both the training
  and validation set, i.e. there is no overfitting,
  even in the Assemblage dataset, which was not part of the training set. 

  SOK's \ArmISA{} results are surprising because \tool{}'s 
  recall in the training set is much lower than in the validation set ($99.188\%$ compared to $99.973\%$).
  Upon closer examination, we noticed this is due to a single binary \texttt{perlbench\_base.\ArmISA{}-gcc81-Os} that has $150,849$ false negatives (over $99\%$ of all the
  false negatives). After excluding it, the resulting overall recall is $99.993\%$.

  Despite their similar performance, learned weights set $46$ (out of $95$) heuristics to $0$. Four of those
  are heuristics reliant on symbols, which produce no matches on stripped binaries. This leaves
  a total of $42$ heuristics ($>40\%$ of all heuristics) that can be removed without a performance impact. 
  Fewer heuristics simplify
  development and code maintenance, making the weight inference algorithm a powerful tool to evaluate
  heuristics and measure their importance.
  For example, \tool{} has two heuristics that assign points when a candidate block is called or jumped to
  by another candidate block respectively. When learning optimal weights, the heuristic that matches jumps gets zero weight
  whereas the one that detects calls has a non-zero weight. 
  This agrees with Priyadarshan et al.'s findings that spurious (short) jumps are much more common than spurious calls in x64~\cite{accuratedisassembly}.

\subsection{Comparison to Other Disassemblers}
\label{sec:comp-orig}
We evaluate \tool{}'s disassembly effectiveness by comparing it to other disassemblers.
SOK's dataset contains disassembly information for several state-of-the-art tools, namely
Binary Ninja, IDAPro, Ghidra, and Angr. We include these results together with ours in Table~\ref{table:other-disassemblers}.
For \tool{}, we use the learned weights
from Sect.~\ref{sec:weight-inference-evaluation}.
In addition, we compare \tool{} to an older version of Ddisasm without interval scheduling (Ddisasm-1.6),
and to two recent disassemblers:
 D-ARM~\cite{d-arm} (commit \texttt{40b5462}) on the \ArmISA{} datasets (Jiang and SOK), 
 and DASSA~\cite{accuratedisassembly} on x64.

\begin{table}[t]
  \caption{Tool comparison: percentage of failures, correctly disassembled binaries, precision, and recall for each dataset and each tool.}
  \label{table:other-disassemblers}

  \begin{tabular}{@{}l@{\,}|@{\,}c@{\,}|r|r|r|r@{}}
    \shortstack{Dataset\\ISA} & Tool & \shortstack{Fail or\\Missing} & Correct & Precision & Recall \\ 
   \hline
     \multirow{7}{*}{ \shortstack{SOK\\x86/x64}} & \tool{} & 0.96\% & 92.8\% & 99.992\% & 99.996\% \\ 
     & Ddisasm-1.6 & 2.44\% &  \textbf{94.9\%} & 99.993\% & \textbf{99.998\%} \\ 
     & DASSA & *49.72\% &  *33.6\% & 99.976\% & 99.948\% \\ 
     & Ninja & 8.76\% &  12.6\% & 99.983\% & 97.151\% \\ 
      & Ida & 0.03\% &  69.1\% & \textbf{99.994\%} & 99.945\% \\ 
      & Ghidra & 20.56\% &  43.0\% & 99.789\% & 93.635\% \\ 
      & Angr & 17.89\% &  60.3\% & 99.908\% & 99.984\% \\

   \hline
     \multirow{6}{*}{ \shortstack{SOK\\\shortstack{\ArmISA{}\\(thumb)}}} & \tool{} & 0.00\% & \textbf{37.3\%} & 99.905\% & \textbf{99.968\%} \\ 
     & Ddisasm-1.6 & 7.09\% &  1.9\% & 99.869\% & 99.848\% \\ 
     & D-ARM & 0.00\% &  0.1\% & 97.733\% & 95.382\% \\ 
     & Ninja & 0.00\% &  0.8\% & 98.577\% & 91.987\% \\ 
      & Ida & 0.00\% &  8.9\% & 99.391\% & 96.388\% \\ 
      & Ghidra & 0.00\% &  1.4\% & \textbf{99.913\%} & 89.046\% \\

      & Angr & 4.14\% &  0.1\% & 97.365\% & 98.682\% \\ 
   \hline
     \multirow{6}{*}{ \shortstack{SOK\\\ArmISA{}}} & \tool{} & 0.08\% & \textbf{51.3\%} & 99.996\% & 99.705\% \\ 
     & Ddisasm-1.6 & 12.72\% &  28.4\% & 99.993\% & \textbf{99.765\%} \\  
     & D-ARM & 0.00\% &  0.0\% & 92.582\% & 91.927\% \\ 
     & Ninja & 0.00\% &  21.7\% & 99.993\% & 96.551\% \\ 
      & Ida & 0.00\% &  11.5\% & \textbf{99.998\%} & 97.955\% \\ 
      & Ghidra & 0.00\% &  1.1\% & 99.969\% & 94.681\% \\

      & Angr & 4.37\% &  15.9\% & 98.524\% & 99.313\% \\ 
   \hline
     \multirow{6}{*}{ \shortstack{SOK\\\AarchISA{}}} & \tool{} & 0.00\% & \textbf{99.7\%} & \textbf{100.000\%} & \textbf{100.000\%} \\ 
     & Ddisasm-1.6 & 0.00\% &  \textbf{99.7\%} & \textbf{100.000\%} & \textbf{100.000\%} \\ 
     & Ninja & 1.19\% &  2.9\%    & \textbf{100.000\%} & 97.274\% \\ 
      & Ida & 3.16\% &  0.0\%     & \textbf{100.000\%} & 97.172\% \\ 
      & Ghidra & 1.19\% &  41.1\% & \textbf{100.000\%} & 97.505\% \\ 
      & Angr & 1.66\% &  96.8\%   & \textbf{100.000\%} & \textbf{100.000\%} \\ 
   \hline
     \multirow{3}{*}{ \shortstack{Jiang\\\ArmISA{}}} & \tool{} & 0.00\% & \textbf{77.0\%} & \textbf{99.993\%} & \textbf{99.996\%} \\ 
      & Ddisasm-1.6 & 0.68\% &  17.6\% & 99.985\% & 99.929\% \\ 
      & D-ARM & 0.00\% &  1.2\% & 96.754\% & 92.249\% \\ 
   \end{tabular}
   
   \scriptsize{*DASSA does not support x86 32bit. It correctly disassembles 67.6\% of the x64 binaries.}
\end{table}

\tool{} obtains the highest recall of all the tools on all datasets.
At the same time, it is second best in precision for SOKs' x86/x64 and \ArmISA{} datasets. In all 
cases, \tool{}'s precision is within $0.01\%$ of the highest precision.
\tool{} achieves the highest rates of correct disassembly in all categories, which is a crucial metric
for binary rewriting applications.

Our \tool{} version improves over Ddisasm-1.6  in all the \ArmISA{} metrics (e.g., 99.99\% vs. 99.92\% on the Jiang dataset)
except for SOK \ArmISA{}'s recall, which is lower due to the outlier discussed in the previous experiment (\texttt{perlbench\_base.arm32-gcc81-Os}).
The improvements are more visible when considering the percentage of correct binaries (binaries with perfect disassembly).
However, \tool{} suffers a small performance regression for x86/x64. 
Nevertheless, our version has fewer failures in that dataset as well (0.96\% vs. 2.44\%).

DASSA's evaluation reports different metrics, but their artifact evaluation also collects precision and recall. These
metrics also differ slightly due to how they are computed. Note that we aggregate all true positives, false positives, and false negatives across all binaries
and compute an overall precision and recall, whereas they compute precision and recall for each binary and then compute
the mean. Our approach gives more weight to larger binaries that contain more instructions rather than averaging over binaries of very different size.
Finally, in DASSA's evaluation, DASSA outperforms Ddisasm (the version is not specified) whereas in our evaluation both
\tool{} and Ddisasm-1.6 yield better results than DASSA. The discrepancy is likely due to DASSA's evaluation
removing exception information from binaries, which could affect Ddisasm's accuracy.
Exception information is used (when available) for the candidate generation traversals (see Sect.~\ref{sec:original-traversal})

Despite our best efforts, we could not reproduce D-ARM's reported results~\cite{d-arm}.
In particular, its authors evaluated D-ARM against the AOSP dataset---a subset of Jiang's dataset
consisting of 669 Android libraries. Their reported precision and recall on AOSP are $99.79\%$ and $99.86\%$. 
In contrast, our experiments yield $97.09\%$ and  $92.42\%$ for D-ARM.
Note that \tool{} achieves $99.992\%$  precision and $99.996\%$ recall for AOSP, which is well
above both the reported and measured D-ARM rates.

\subsection{Runtime Performance}
\label{sec:runtime-performance}

We conduct two sets of runtime measurements to evaluate the
performance of \tool{}.  First, in Section~\ref{sec:compare-to-1.6},
we demonstrate the practicality of \tool{}'s extended traversal and
conflict resolution algorithm  by comparing its runtime against
Ddisasm-1.6.  Second, in Section~\ref{sec:runtime-weight-inference},
we assess the overhead introduced by collecting and solving
constraints for weight inference.

\subsubsection{\tool{} vs. Ddisasm-1.6}
\label{sec:compare-to-1.6}

\begin{table}[t]
    \centering
    \caption{\tool{} vs. Ddisasm-1.6 Runtime Comparison: Number of binaries
    that both tools succeeded on, total runtimes,
    and \tool{} runtime relative to Ddisasm-1.6.}
    \begin{tabular}{@{}c@{\,}|r|r|r|r@{}}
    Dataset & \# Common & \multicolumn{2}{c|}{Runtime (s)} & Ratio \\
        ISA & Binaries & \tool{} & Ddisasm-1.6 & WIS/1.6 \\
        \hline
        \shortstack{SOK\\x86/x64} & 3,868 & 165,513 & 191,850 & 86\% \\
        \hline
        \shortstack{SOK\\\ArmISA{} (thumb)} & 1,167 & 37,702 & 33,437 & 112\% \\
        \hline
        \shortstack{SOK\\ \ArmISA{}} & 1,138 & 42,694 & 40,535 & 105\% \\
        \hline
        \shortstack{SOK\\ \AarchISA{}} & 1,264 & 21,458 & 16,735 & 128\% \\
        \hline
        \shortstack{Jiang\\\ArmISA{}} & 1,019 & 20,651 & 16,512 & 125\% \\
        \hline
        \shortstack{Pangine\\x86/x64} & 839 & 137,862 & 124,103 & 111\% \\
    \end{tabular}
    \label{tab:runtime_performance} \\
\end{table}

Table~\ref{tab:runtime_performance} shows the comparison of the total
runtimes of \tool{} against Ddisasm-1.6.  In all datasets except
SOK x64/x64, \tool{} is slightly slower than Ddisasm-1.6 (at most 28\% slower
in the worst performing category \AarchISA).
While the comparison does not isolate all the factors that determine this
performance difference, it demonstrates that \tool{}'s improved disassembly algorithm
is practical. Given the improved results of our tool, we
believe that this runtime increase is acceptable.

\subsubsection{Weight Inference}
\label{sec:runtime-weight-inference}

\begin{table}[t]
    \centering
    \caption{Constraint Inference Runtime: Number of binaries,
    total contraint inference (CI) runtime, disassembly runtime (D),
    and CI runtime relative to disassembly.}
    \begin{tabular}{@{}c|c|c|c|c@{}}
       Dataset &  & \multicolumn{2}{c|}{Runtime (s)} & Ratio\\
       ISA & \# Binaries & CI & D & CI / D \\
        \hline
        \shortstack{SOK\\x86/x64} & 738 & 4,909 & 22,542 & 21\% \\
        \hline
        \shortstack{SOK\\\ArmISA{} (thumb)} & 260 & 1,440 & 7,205 & 19\% \\
        \hline
        \shortstack{SOK\\\ArmISA{}} & 274 & 13,284 & 8,509 & 156\% \\
        \hline
        \shortstack{SOK\\\AarchISA{}} & 253 & 572 & 3,746 & 15\% \\
        \hline
        \shortstack{Jiang\\\ArmISA{}} & 199 & 2,340 & 4,661 & 50\% \\
        \hline
        \shortstack{Pangine\\x86/x64} & 171 & 5,918 & 15,701 & 37\% \\
    \end{tabular}
    \label{tab:runtime-weight-inference} \\
\end{table}

The results in Table~\ref{tab:runtime-weight-inference} highlight  the runtime overhead introduced by collecting  weight constraints, offering insight into the efficiency of the weight-inference step in our algorithm.
In five out of six datasets, excluding SOK \ArmISA{}, weight inference completes significantly faster than the overall disassembly process, often taking less than 50\% of the time total runtime.
The only outlier is the SOK \ArmISA{} dataset, where weight inference takes longer than disassembly.
In this dataset, the disassembly algorithm creates large candidate data blocks representing jump tables, which prevents effective interval scheduling decomposition (see Sect.~\ref{sec:scheduling-decomposition}), thereby harming the constraint inference performance. 
Nonetheless, since weight inference is a one-time preprocessing step whose results are leveraged by \tool{}, we argue that its overhead remains acceptable.
Pulp solved the LP problem with the 86,263 collected constraints in approximately 10 seconds.

\section{Related Work}
\label{sec:related}
Our work focuses on static disassembly and, in particular, on instruction recovery.
It extends Ddisasm~\cite{ddisasm} with an 
extended candidate block generation to handle \ArmISA{} and \AarchISA{} binaries,
and a novel conflict resolution algorithm. 
In contrast to the original Ddisasm, which only considered overlaps between two
blocks at a time, our conflict resolution algorithm is expressed as a
weighted interval scheduling (WIS) problem that jointly considers all block overlaps.

D-ARM's~\cite{d-arm} is a closely related approach.
We note three important differences.
First, D-ARM is tailored for arm and \AarchISA{} binaries, whereas our approach performs well for
x86, and x64 as well.
Second, D-ARM does not have a candidate generation phase. D-ARM instead considers all individual
instructions (superset disassembly) as candidates.
Third, D-ARM's conflict resolution is encoded as a maximum weight independent set (MWIS)
optimization problem. While WIS only considers overlaps, MWIS
can enforce control flow dependencies as well. 
For example, if a candidate code block $A$ has a call to another candidate code block $B$,
then $A$ implies $B$.
These dependencies make the optimization problem NP-Hard, which
dictates the adoption of greedy approximation methods.

As we have argued in Sect.~\ref{sec:algorithm}, grouping instructions into blocks
is especially important for x86 and x64, because it reduces the number of block candidates by an order of 
magnitude compared to superset instructions.
We believe that our candidate generation phase and control flow-based heuristics compensate for the lack of dependencies 
in the conflict resolution phase. Considering multi-instruction candidate blocks effectively enforces dependencies
among instructions in the same code block.
Our experiments indicate that WIS is sufficient, given an appropriate 
set of heuristics. However, further research should be conducted to understand the tradeoffs
of both approaches better.

Priyadarshan et al.~\cite{accuratedisassembly} recently presented the DASSA disassembler.
DASSA is focused on x64. It computes a superset of all
possible code (akin to our candidate generation) and also implements a conflict resolution method. 
DASSA's conflict resolution is greedy. Each step considers a single candidate with
the highest score (weight).
For each candidate, DASSA computes a score based on statistical properties of the data
(DASSA refines the probabilistic analysis of previous works~\cite{probabilistic}), and 
several static analysis-based checks to flag invalid code.
Whether a candidate is selected depends on a combination of both the score and the checks.
In contrast, our approach encodes both statistical properties and static analysis-based checks 
into a single weight. This results in a simpler conflict resolution algorithm. This integration is 
facilitated by our weight inference algorithm,
which automatically adjusts the relative weights of the different heuristics.
Integrating DASSA's analyses and checks into our approach could further improve its accuracy.

Recent works attempt to leverage deep learning for disassembly, notably XDA~\cite{pei2021xda}
and DeepDi~\cite{DeepDi}. XDA analyzes raw bytes directly using a specially designed language model. DeepDi
performs superset disassembly~\cite{superset}, and embeds instruction using a graph neural network that captures
dependencies among instructions. Both D-ARM and DASSA's evaluations compare against and outperform XDA.
DeepDi's evaluation shows that Ddisasm (the version is not specified) achieves higher accuracy than DeepDi.


Traditional disassemblers also combine static analyses and heuristics. Pang et al.~\cite{sok-x86} provides
a detailed overview for x86/x64 and Jiang et al.~\cite{arm-evaluation} perform an extensive comparative of \ArmISA{} disassemblers.

Most static disassemblers \cite{idapro,ghidra,hopper,BAP,ddisasm,binja,radare,angr} recover
additional information, such as control flow graphs, function boundaries~\cite{nucleous},
symbolization information~\cite{ramblr,RetroWrite,uroboros},
or even perform decompilation~\cite{decompilation}. Disassembly can also be performed dynamically,
or through a combination of static and dynamic techniques~\cite{Zhang2021StochFuzzSA}.
Despite our significant progress (see Table~\ref{table:other-disassemblers}), 
binary rewriting applications have very low tolerance for errors. Thus, multiple approaches have been proposed
to instrument programs without requiring perfect disassembly \cite{superset,E9Patch,armmore}, usually
at the expense of some performance overhead.


\section{Conclusion}

In this paper we present a novel three-phase disassembly algorithm: (1) a candidate generation phase produces
an overapproximation of all code blocks; (2) a weight assignment phase assigns a weight to each candidate block; and (3) a conflict
resolution phase resolves overlaps producing the final set of blocks.
Conflict resolution is expressed as a WIS problem, which has an efficient optimal solution.

We also present a solution to the weight assignment problem, i.e. how to optimally assign weights to heuristics to maximize
accuracy. Our solution optimizes heuristic weights based on binaries annotated with ground truth. For each binary, it
collects a set of linear constraints over unknown heuristic weights that ensure correct disassembly. 
We maximize the number of satisfied constraints from all the binaries by solving a LP
problem with soft constraints.

We have implemented our approach on top of Ddisasm~\cite{ddisasm} and performed a large experimental evaluation to validate it. 
Our approach supports multiple ISAs (x64, x86, \AarchISA{}, and \ArmISA{}) and outperforms
 (in number of correct binaries and recall)
 or nearly matches (with less than 0.002\% difference in precision) SOTA disassemblers.
Our weight inference algorithm results in weights that outperform manually tuned weights in several benchmarks, while
effectively reducing the number of heuristics (by setting their weights to 0) by 40\%.

\bibliographystyle{plain}
\bibliography{biblio}

\appendices

\section{Proofs}
\label{app:proofs}

This section contains the proofs for Theorem~\ref{theorem:sound-overapproximation}
and Theorem~\ref{theorem:infer-constraint-correctness}.
For convenience, we restate the theorems before proceeding with the proof.

\setcounter{theorem}{0}

\begin{theorem}
    If a candidate block set $\CandidateBlocks$ does not present missed instructions nor incorrect
    code block boundaries, then $\CandidateBlocks$ is a sound overapproximation.
\end{theorem}

\begin{proof}[Proof of Theorem~\ref{theorem:sound-overapproximation}]
We prove that we can build a subset $\BlockSet\in\CandidateBlocks$ such that $\bigcup_{b\in \BlockSet} \BlockInsn(b) = \Code$.
Our proof relies of the fact that candidate code 
blocks do not share instructions (See Sect.~\ref{sec:original-traversal}).
If $\CandidateBlocks$ has no missed instructions, every $i\in \Code$ belongs to
a single block $b\in \CandidateBlocks$ and $\BlockOf(i)$ (see Definition~\ref{def:code-block}) yields a single block for each $i\in \Code$. 
Thus, we can define our final block set
$\BlockSet = \bigcup_{i\in \Code} \BlockOf(i)$. By construction $\bigcup_{b\in \BlockSet} \BlockInsn(b) \supseteq \Code$.

Next, we prove that $\bigcup_{b\in \BlockSet} \BlockInsn(b) \subseteq \Code$, which implies that 
$\bigcup_{b\in \BlockSet} \BlockInsn(b) = \Code$ and consequently $\CandidateBlocks$ is a sound overapproximation.
If $\CandidateBlocks$ has no incorrect code block boundaries, for every block $b\in\CandidateBlocks$ we have 
either $\BlockInsn(b) \cap \Code = \emptyset$ or $\BlockInsn(b)\subseteq \Code$.
Since we have defined our set $\BlockSet$ using $\BlockOf$, we know the first condition does not hold.
Therefore, for every $b\in \BlockSet$ we know that $\BlockInsn(b)\subseteq \Code$, which implies
$\bigcup_{b\in \BlockSet} \BlockInsn(b) \subseteq \Code$.
\end{proof}

\begin{theorem}
    Let $\constraintSet$ be the set of constraints generated by Algorithm~\ref{alg:wis-constraint-inference-naive} and
    let $\alpha: H \rightarrow \mathbb{Z}$  be a weight assignment that satisfies the constraints $\alpha \models \constraintSet$, 
  then there is an optimal schedule $S$  such that $\alpha(\sweight(S))> \alpha(\sweight(S'))$ for every non-optimal schedule $S'$.
  \end{theorem}

\begin{proof}[Proof of Theorem~\ref{theorem:infer-constraint-correctness}]
First, we prove that at the end of the loop,
 $\symopt[n]$ contains tuples representing all possible schedules with non-overlapping blocks.
    The same argument that ensures the correctness of WIS algorithm (Algorithm~\ref{alg:weightedinterval})
    guarantees can be applied here.
    
    We prove it by complete induction over $i$. 
    Our inductive invariant is that at each iteration $i$, $\symopt[i]$ contains tuples for all schedules including
    blocks from $b_1$ to $b_i$. The base case (i=0) is trivial with an empty schedule, and the inductive step $i$ includes
    all possible schedules with $b_i$ ($\mathit{take}$) and all the possible schedules without $b_i$ ($\mathit{leave}$).
    
    Since $\symopt[n]$ contains all the possible schedules, $\bestSchedule$ will return an optimal schedule $S$.
    Line~\ref{line:non-optimal} collects all non-optimal schedules by relying on Lemma~\ref{lemma:optimal-score}, and Line~\ref{line:cs} generates a constraint
    $\sweight(S) > \sweight(S')$ for each non-optimal schedule. Thus, for each non-optimal schedule
    $S'$ there is a constraint $\sweight(S) > \sweight(S')$ in $\constraintSet$. Since $\alpha \models \constraintSet$
    it satisfies $\alpha \models \sweight(S) > \sweight(S')$ as well.
\end{proof}
\section{Ground Truth Limitations}
\label{app:gt-limitations}

\begin{table*}[t!]
    \centering
    \caption{Manual Assessment of Unsound PE Binaries}
    \begin{tabular}{|l|l|p{3.9cm}|p{3.9cm}|}
        \hline
        \textbf{Category} & \textbf{Affected Binaries} & \textbf{Issue Description} & \textbf{Impact on Disassembly} \\
        \hline
        Missing Jump-Table Information & 
        \begin{tabular}[t]{@{}l@{}}SymbolLoadDLL\_d.dll, ChisMath.exe, \\ DrawDemo.exe, GetScreenRGB.exe, \\ osu-memory-demo.exe, threadpool.exe\end{tabular} & 
        PDB files lack information about jump tables, leading to incomplete or incorrect function boundaries. & 
        Our tool correctly identifies jump tables, leading to discrepancies compared to PDB ground truth. \\
        \hline
        Misinterpreted Padding Bytes & 
        \begin{tabular}[t]{@{}l@{}}MaiSense.dll, iphubclient\_amxx.dll, \\ gtest\_unittest.exe, SysInv32.exe, \\ ProfilerOBJ.dll, Test.exe\end{tabular} & 
        A \lstinline!jmp! instruction targets an sequence of \lstinline!int 3! instructions, typically used as padding. & 
        Our tool treats \lstinline!int 3! sequences as padding rather than executable code, marking
        the \lstinline!jmp! instruction targeting them as data as well. \\
        \hline
        Obfuscated Code & 
        \begin{tabular}[t]{@{}l@{}}hovew-crackme.exe, SPASM.exe\end{tabular} & 
        Unusual control-flow structures and inline data embedded within code segments result in incorrect PDB information. & 
        PDB files misrepresent function boundaries, while our tool correctly handles the obfuscated code. \\
        \hline
        \tool{} bug & 
        \begin{tabular}[t]{@{}l@{}}psp.exe\end{tabular} & 
          A minor bug causes a call instruction target to be computed incorrectly, which results in the
          instruction invalidation. & 
        Future work will address this case. \\
        \hline
    \end{tabular}
    \label{tab:pdb_issues}
\end{table*}

During our experiments, we have identified multiple limitations in the ground truth provided by the different
dataset (see Table~\ref{table:datasets}). In this section, we detail our findings and how we address them in our
experiments.

\subsection{SOK ground truth limitations}
SOK's ground truth does not distinguish ARM and Thumb decode modes, instead it has two separate datasets
for each mode. Unfortunately, we have observed that some Thumb binaries contain a few ARM functions, usually added
by the linker. SOK's evaluation tries to exclude functions added by the linker (we add those to $\Ignored$), but it
fails if the binaries are stripped before collecting the ground truth \footnote{\anonymizedUrl{https://github.com/junxzm1990/x86-sok/issues/32}}.
Binaries can be recompiled to avoid stripped binaries, but that would also require rerunning all the disassemblers
on the newly compiled binaries, since small changes in the compilation environment could lead to differences 
in the generated binaries.

Instead, we reuse the originally provided datasets and we detect ARM functions in Thumb binaries 
by checking the consistency of the reported instruction sizes. If the
ground truth reports an instruction size of $4$ at address and the instruction decoder returns
a instruction of size $2$ at that same address using Thumb mode, then that instruction (and its corresponding function)
must be in ARM decode mode. Note that naively looking for $4$ byte instructions is not enough because some
some Thumb instructions are indeed $4$ bytes long. Note also that this method might not find all ARM functions in Thumb
binaries, which could introduce some small errors in the \ArmISA{} Thumb evaluation.

During our experiments, we have also identified other errors in SOK's ground truth,
which we have reported\footnote{\anonymizedUrl{https://github.com/junxzm1990/x86-sok/issues/31}}.
These errors were cause by (1) a failure to correctly handle duplicated sections during linking,
(2) an unsupported instruction \texttt{udf} in the compiler instrumentation that recovers the ground truth.
Both issues have been promptly resolved by the authors, but the datasets have not been re-generated.
We exclude manually excluded the detected cases of (1) and extended our evaluation scripts to recover the
missing \texttt{udf} from the ground truth.

\subsection{Pangine ground truth limitation}

This ground truth is limited to regions within well-defined function boundaries, and thus it is incomplete. We add the regions outside  function boundaries to $\Ignored$.
In addition, we have also found and reported issues with Panguine's ground truth\footnote{\anonymizedUrl{https://github.com/pangine/disasm-benchmark/issues/2}}.
Some MSVC-compiled binaries contain jump tables being classified as code. These are classified as \emph{optional} true instructions,
which indicates a lower degree of certainty in the ground truth. We followed the authors' recommendation and ignored optional
true instructions in our evaluation scripts.

\subsection{Jiang ground truth limitations}

Jiang's \ArmISA{} dataset relies on compiler-generated marker symbols:
\$a, \$t, and \$d for ARM, Thumb, and data respectively for extracting ground truth information.

Unfortunately, marker symbols are not always accurate on binary regions added by the linker.
We account for this limitation by considering marker symbols only within the boundaries of 
function symbols (other regions are added to $\Ignored$). We have found this restriction to yield more reliable ground truth
at the expense of completeness (larger parts of the binaries are ignored). 
In addition, we have also semi-automatically annotated some of those ignored regions that contain
specific patterns, such as ARM/Thumb interworking veneers, that can be matched with high confidence.
Finally, we have completely excluded $7$ binaries where we found incorrect ground truth.
These are mostly binaries implementing cryptographic primitives (such as \texttt{libcrypto})
where developers have encoded instructions as data directly (resulting in incorrect marker symbols).

\subsection{Assemblage ground truth limitations}

While PDB files are a useful source of ground truth for evaluating disassembly, they are not entirely reliable, particularly in cases involving jump tables and obfuscation. See Appendix~\ref{app:assemblage-pdb} for more details.

\section{Manual Assessment of Unsound PE Binaries}
\label{app:assemblage-pdb}

Our candidate generation evaluation
(Sect.~\ref{sec:experiments:candidates}), results in a significantly higher ratio
of unsound binaries for the Assemblage PE binaries (0.61\% vs. 0.35\%
for second worst category).  To investigate further, we randomly
selected 15 binaries from the 518 unsound binaries (Table~\ref{table:candidates}) and manually inspected
them. The results of this analysis can be found in Table~\ref{tab:pdb_issues}. We find
three primary causes for binaries being classified as unsound: missing jump-table
information in PDB files, misinterpretation of padding bytes, and
obfuscated code \footnote{Given the limited size of the sample, additional types of issues may exist, and the relative frequency of each category may differ in the full dataset.}.

\subsection{Missing Jump-Table Information}
We found six binaries with missing jump-table information in the
PDB debug data.  Jump tables are commonly used for indirect branching,
typically within the code section of PE binaries.  Our tool
successfully reconstructed these jump tables, correctly identifying
function boundaries and indirect branch targets.  The discrepancy
between our generated candidates and the PDB ground truth is,
therefore, due to a limitation of the PDB files.

\subsection{Misinterpreted Padding Bytes}
MSVC binaries use \lstinline{int 3} (software breakpoint) instructions
as padding (to align function boundaries).  
While having a single \lstinline{int 3} instruction as part
of the executable code is possible, a sequence of \emph{consecutive} \lstinline{int 3}
instructions is very unlikely to be meant for execution. In other words, sequences
of \lstinline{int 3} instructions are indeed very common in PE binaries, but they are never executed. Thus, \tool{}
considers sequences of \lstinline{int 3} as invalid. This invalidation is then 
propagated backwards to instructions jumping to those sequences directly during
the first disassembly traversal (see backward traversal in
Section~\ref{sec:original-traversal}). 
We find instructions in the PDB
ground truth being invalidated because of this reason in six
binaries. We believe these jumps to \lstinline{int 3} sequences might be
a case of obfuscation. Unfortunately, the PDB ground truth does not
explicitly clarify whether such \lstinline{int 3} sequences should be
considered executable (since the PDB ground truth is incomplete),
making it challenging to establish a definitive interpretation.

\subsection{Obfuscated Code}
In two instances, we found obfuscated code patterns designed to confuse disassembly tools. For example, we found conditional jumps targeting the middle of multi-byte instructions or raw byte sequences (not decodable as instructions), which result in ambiguous control flow and overlapping instruction boundaries. Other examples included manually inserted bytes that disrupted typical linear disassembly.
In these cases, the PDB files provided incorrect instruction boundaries, whereas our tool, leveraging more heuristics, correctly disassembled the obfuscated code.

\subsection{Other Cases}
The remaining unsound binary, \texttt{psp.exe},  is caused by a bug in our tool.  We plan to address this bug shortly.

\end{document}